\documentclass[a4paper,12pt]{article}
\usepackage[utf8]{in
    putenc}
\usepackage{cancel}
\usepackage{ulem}
\usepackage{amsfonts}
\usepackage{amssymb}
\usepackage{multicol}
\usepackage{graphicx}
\usepackage{amsmath,bm}
\usepackage{enumerate}
\usepackage{mathtools}
\usepackage{tikz} \usetikzlibrary{calc}
\usepackage[countmax]{subfloat}
\usetikzlibrary{fit}
\usepackage{tabularray}
\usepackage{longtable}
\usepackage{xcolor}
\usepackage{float}
\usepackage{color}
\usepackage{here}
\usepackage{cite}
\usepackage{tikz-3dplot}
\usepackage{subcaption}
\usepackage{mwe}
\usepackage[outline]{contour}
\usepackage{tabularx}
\usepackage{multirow}
\usetikzlibrary{3d}
\usepackage{mathrsfs}
\usepackage{float,epsfig}
\usepackage{dcolumn}% Align table columns on decimal point
\usepackage{pgfplots}
\usepackage{array}
\pgfplotsset{compat=1.3}
\usepackage{graphicx}% Include figure files
\usepackage{bm}% bold math
\usepackage{amsmath,amssymb,amsthm}
\usepackage[colorlinks=true,linkcolor=blue,citecolor=red]{hyperref}
\DeclareSymbolFont{symbols} {OMS}{cmsy}{m}{n}

\def\be{\begin{equation}}
\def\ee{\end{equation}}
\def\bea{\begin{eqnarray}}
\def\eea{\end{eqnarray}}
\def\ba{\begin{aligned}}
\def\ea{\end{aligned}}

\def\pp{\partial}

\usepackage{mathtools}
\newcommand\myeq{\mathrel{\stackrel{\makebox[0pt]{\mbox{\normalfont\tiny $\tilde{R}\rightarrow\infty$}}}{=}}}
\newcommand\myeqq{\mathrel{\stackrel{\makebox[0pt]{\mbox{\normalfont\tiny $\tilde{R}\rightarrow\infty$}}}{\rightarrow}}}

\usepackage{booktabs}

\newcommand\tablex{2.5mm}
\newcommand\tabley{1mm}

\definecolor{lime}{HTML}{A6CE39}
\newcommand{\orcidicon}{%
    \begin{tikzpicture}
    \draw[lime, fill=lime] (0,0)
        circle [radius=0.16]
        node[white] {{\fontfamily{qag}\selectfont \tiny ID}};
    \draw[white, fill=white] (-0.0625,0.095)
        circle [radius=0.007];
    \end{tikzpicture}   \hspace{-2mm}
}
\newcommand\orcidHasan{{\href{https://orcid.org/0000-0001-7408-0910}{\orcidicon}}}
\newcommand\orcidKarima{{\href{https://orcid.org/0000-0001-5419-8516}{\orcidicon}}}

\newcommand\orcidLahoucine{{\href{https://orcid.org/0000-0002-0143-5140}{\orcidicon}}}
\title{\bf On M87$^*$ and SgrA$^*$ Observational Constraints  of Dunkl Black Holes  

}

\author{ N. Askour$^1$,  A. Belhaj$^2$,  L. Chakhchi$^3$\orcidLahoucine\!\!\thanks{lahoucine.chakhchi@edu.uiz.ac.ma (Corresponding author)}, H.   El Moumni$^3$\orcidHasan\!\!, K. Masmar$^3$\orcidKarima\!\!\footnote{ Authors in alphabetical order} 
\\
{\small $^{1}$Department of Mathematics, Sultan Moulay Slimane University, Faculty of Sciences and Technics}
\\
{\small  Béni Mellal, BP 523, 23000, Morocco.}
\\
{\small $^{2}$Physics Department, Faculty of Science, Mohammed V University in Rabat, Rabat, Morocco.}
\\
{\small $^{3}$LPTHE, Physics Department, Faculty of Science,  Ibnou Zohr University, Agadir, Morocco.}
}

\vspace{-3.6em}

\date{}
\begin{document} 
\maketitle
\begin{abstract}

In this work, we  investigate   the optical properties  of a new black hole recently obtained  from the Dunkl operator formalism involving a relevant parameter  denoted by $\xi$.  Concretely,   we  first investigate   the shadows,  the Lyapunov exponents of unstable nearly bound
orbits and  the  spherically infalling accretion behaviors in terms of such a parameter.  Then,  we examine the effect of  this parameter  on
the  Dunkl  black hole deflection angle in vacuum and medium backgrounds by manipulating   the Gauss-Bonnet theorem. 
 Exploiting  the  M87$^*$ and SgrA$^*$  optical bonds,   we  provide strong  constraints  on  $\xi$ via  the falsification mechanism.

%\newline\newline	  
%{\it Keywords}: Dunkl Operator; Gauge Theory; Thermodynamic Properties.

%{\noindent}

\end{abstract}

\newpage
\tableofcontents

%\newpage

\section{Introduction}
Recently, the  physics of black holes has attracted a great deal of interest via various  theoretical  and  empirical  investigations  including the Even Horizon Telescope (EHT) collaborations and  the detection of gravitational  waves \cite{EventHorizonTelescope:2019dse,EventHorizonTelescope:2022wkp,Abbott:2016blz}. Theoretically, the thermodynamics\cite{Chamblin:1999tk,Belhaj:2012bg,Gunasekaran:2012dq,Hendi:2012um,Zhang:2015ova,Wei:2015iwa,Perlick:2018iye,Nguyen:2015wfa,Kubiznak:2012wp,Belhaj:2015hha,Chabab:2015ytz,Chabab:2016cem,Chabab:2019kfs,Chabab:2018lzf,Zou:2017juz} and  the optic\cite{Gralla:2019xty,Bambi:2013nla,Narayan:2019imo,Falcke:1999pj} are the most studied subjects by considering gravity theories  in various geometric   backgrounds such as  the cosmological spacetimes and the Universe  dark sectors.  In connection with  the phase transitions, the anti de Sitter (AdS)  spaces have been placed in the center of such investigations \cite{Barzi:2023msl,Ali:2023jox,Chabab:2019kfs}.  Transforming the cosmological constant into pressure, several phase transitions, such as the Hawking-Page one,  have been  investigated  in the presence of ordinary and non-ordinary parameters. Different scenarios, including  the criticality and  the stability  properties,   have been  explored to inspect the thermodynamic behaviors of the AdS  black holes. Various impacts as  the spacetime dimension,  the dark matter and  the dark  energy have been extensively studied over the last few years.  It has been found    certain  significant modifications  depending  on   the extra parameters in arbitrary  dimensions.

With reference to  the  EHT collaborations, the light behaviors near  black holes have been largely investigated. The shadow configurations,   considered as the  indirect test of the even horizon,  have  been approached by considering different gravity  theories  including the  modified ones \cite{Nojiri:2017ncd,Shankaranarayanan:2022wbx}.   These activities are supported and encouraged by the imaging of  the supermassive    M87$^*$   black hole. This has been confirmed  by the image of the SgrA$^*$  black hole.  These  observational findings  have shown  the real behaviors of the underlying physics of black holes \cite{EventHorizonTelescope:2019pgp,EventHorizonTelescope:2019ggy,EventHorizonTelescope:2022exc,EventHorizonTelescope:2022urf,EventHorizonTelescope:2022xqj}. Motivated by such an  empirical aspect,  the shadows of several black holes have been elaborated. Concretely, different sizes and   shapes have been obtained  by combing  the   general relativity (GR) and  the hamiltonian formalism techniques. Precisely, the photons near  black holes  have been approached by  establishing the associated  equations of motion     via numerical  computation methods. Circles and deformed  ones  have been obtained depending on   the absence or  the presence of certain black hole parameters\cite{Takahashi:2004xh,Wei:2019pjf,Grenzebach:2015oea,Zhang:2020xub,Chakhchi:2024tzo}.  The size and  the shape deformations have been examined by help of 
the geometrical observables using  appropriate numerical simulations. Alternatively,  the deflection of light rays in curved geometries have been also investigated in connection with the  black hole  physics.  Considering a static spherically symmetric solutions,  the light deflection angles  by various black holes  have  been computed and inspected in the presence  of  non trivial parameters.  The Gauss-Bonnet theorem  is considered as  the most used  method to provide interesting results on the deflection angle \cite{Ishihara:2016vdc,Crisnejo:2018uyn,Ovgun:2018oxk,Jusufi:2018kry,Aebischer:2019mtr,Ovgun:2020yuv,Belhaj:2022vte}.  This  computing  road, proposed first  by  Gibbons and Werner \cite{Gibbons:2008rj,Werner:2012rc}  in the context of the optical geometries for asymptotically
flat  backgrounds, 
 has been  exploited  to explore  a bridge to thermodynamics \cite{Walia:2021emv,Belhaj:2021tfc}. In particular, it has been shown that this quantity could provide data on the phase transitions in the  AdS black holes.   Based on the elliptic functions,   the phase structure of  the charged AdS  black  holes  has  been studied  from    thermal variations of the deflection angle.  Precisely, it has been suggested  that  the large black hole/small black hole transition  can  occur naturally at  specific  values of the  deflection angle.

More recently, the empirical activities provided by the  EHT collaborations have been exploited to show the validity and the  viability of certain modeled black holes including the non-Kerr solutions.  Concretely,  the  M87$^*$ and  the SgrA$^*$ black hole measurements  could provide strong conditions  for the proposed models.  More precisely,  the EHT bounds on the shadows  have been investigated to put constraints on the black hole parameters including the charge \cite{EventHorizonTelescope:2021dqv,Zakharov:2021gbg,EventHorizonTelescope:2022xqj,Chakhchi:2024tzo,Chakhchi:2024obi}.

In relation to  the EHT  observational  findings,  moreover,  certain constraints on the  deflection angle  in the dark matter backgrounds  have been studied  in order to unveil information on  the underlying  interactions. Among others,  it has been shown  that the  deflection angle  can supply    an alternative  road to  detect  
the dark matter  going beyond the black hole shadow investigations \cite{Pantig:2022toh}. 

Motivated  by such activities, we investigate  the optical behaviors of a new black hole recently obtained in \cite{Sedaghatnia:2023fod}. The latter will be called Dunkl black hole derived from the implementation of  the Dunkl operator formalism  in the gravity calculations. The  resulting model involves  a relevant parameter  denoted by $\zeta$ being the  center of the present investigation.  Precisely,  we first   discuss    the shadow,  the Lyapunov exponents of unstable nearly bound
orbits and  the  spherically infalling accretion behaviors in terms of such a parameter.  After that, we  study 
the  Dunkl  black hole deflection angle in vacuum and medium backgrounds by manipulating   the Gauss-Bonnet theorem. 
Employing  the  M87$^*$ and SgrA$^*$  optical bonds,   we  supply  strong  constraints  on  $\xi$ via  the falsification mechanism.

The organization of this paper is as follows. In section 2, we  reconsider the study  of the Dunkl black hole. Section 3  fournishes the computations  dealing with  the shadows and the deflection angle.  In section  4, we exploit the EHT bands to determine  certain constraints of the Dunkl parameter $\xi$ from   optical  computations.
In this  work,  the natural units $G = c = 1$  have been used.

\section{Dunkl black hole}

\begin{figure}[!ht]
\centering 
\includegraphics[width=1.1\textwidth]{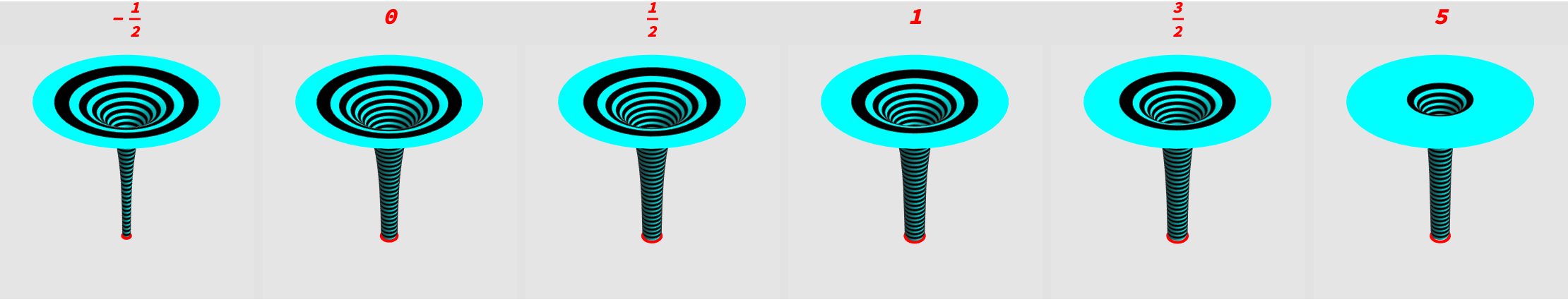}
\caption{\label{fig1} \footnotesize   \it
Embedding diagram within different values of the correction parameter $\zeta$, with  $M=1$.
}
\label{plot1}
\end{figure}

In this section, we reconsider the study of  the Dunkl black hole recently proposed in \cite{Sedaghatnia:2023fod}.  In particular,  we provide the essentials on the associated  solutions from the Dunkl operator formalism combined with gravity computations.  For simplicity reasons, we   deal with   the spherical solution via the  following line element    
\begin{eqnarray}\label{22d}
		ds^{2}=-f(r) dt^{2}+\frac{1}{f(r)}dr^{2}+r^{2}\left(d\theta^{2}+\sin^{2} \theta d \phi^{2}\right)
		\end{eqnarray}
where  the  metric function  $f(r)$  carries physical data on the black holes. The coefficients of such a  radial  function   parametrizes   a space called the moduli space.  Certain regions of such a space can  provide results that could be 
corroborated by   EHT empirical findings.  In the present work, the  moduli space  will  be approached  via   a new  relevant parameter derived naturally  from  the Dunkl operator formalism being largely investigated in connection with various mathematical and physical subjects \cite{dunkel,DUNKL1999819,Chung:2021jfd,Salazar-Ramirez:2016bpi}. It has been remarked that   the Dunkl derivative is a generalization of the ordinary derivative in  the Euclidean spaces involving  reflection  operations.  These  generalised derivatives  introduced by C. F. Dunkl  rely on  reflection symmetries associated with  non-trivial  mathematical structures including  the Coxeter groups and  the  root systems \cite{dunkel,askour}. In Cartesian  geometry, the  general expression of the Dunkl operators   take the following form
\begin{eqnarray}\label{4}
D_{x_{i}}= \dfrac{\partial}{\partial x_{i}}+\dfrac{\alpha_{i}}{x_{i}}(1-\mathcal{R}_{i})\quad,\quad i=0,1,2,3
\end{eqnarray}
	where  one has used  $\alpha_{i}=\left(0,\alpha_{1},\alpha_{2},\alpha_{3}\right)$ $(\alpha_{i}> -1/2)$ being the Dunkl parameters.  $\mathcal{R}_{i}=\left(0,\mathcal{R}_{1},\mathcal{R}_{2},\mathcal{R}_{3}\right)$   denote  the parity operators\cite{DUNKL1999819,Chung:2021jfd,Salazar-Ramirez:2016bpi}. 
 Combining  the Einstein equation and the  Dunkl operator formalism,  a deformed Schwarzschild spacetime resulting from the gauge theory of gravity in the presence of  cosmological constant  has been constructed  \cite{Sedaghatnia:2023fod}.  For  a vanishing  cosmological constant scenario, the metric function  $f(r)$  has been found to be 
\begin{eqnarray}\label{2}
	f(r)=\frac{1}{\left(1+\zeta\right)}-2 M\, r^{\frac{1}{2}\left(1-\sqrt{9+8 \zeta}\right)}\,,%-\frac{\Lambda}{3}\,r^{\frac{1}{2}\left(1+\sqrt{9+8 \zeta}\right)}. \quad
	\end{eqnarray}
Here,  $M$ denotes the black hole mass, %$\Lambda$ denotes the cosmological constant 
and   $\zeta$  is 
the Dunkl parameter 	where one has used  $\zeta=\sum_{i=1}^3 \alpha_i\left(1-\mathcal{R}_{i}\right)\left(1+\alpha_i\left(1-\mathcal{R}_{i}\right)\right)$. This parameter  being relevant in the present investigation enlarges the ordinary  moduli space of  the neutral %AdS 
black holes without  rotations.  It has been revealed that the associated geometrical aspects have been corrected by such a  Dunkl parameter $\zeta$.  
To disclose some pieces of information about the Dunkl parameter $\zeta$ impact,  the embedding diagram of such a black hole within different values of the Dunkl   parameter $\zeta$  is  illustrated in  Fig.\ref{plot1}.

\begin{figure}[!ht]
\centering 
			\begin{tabbing}
			\centering
			\hspace{9.cm}\=\kill
\includegraphics[width=.44\textwidth]{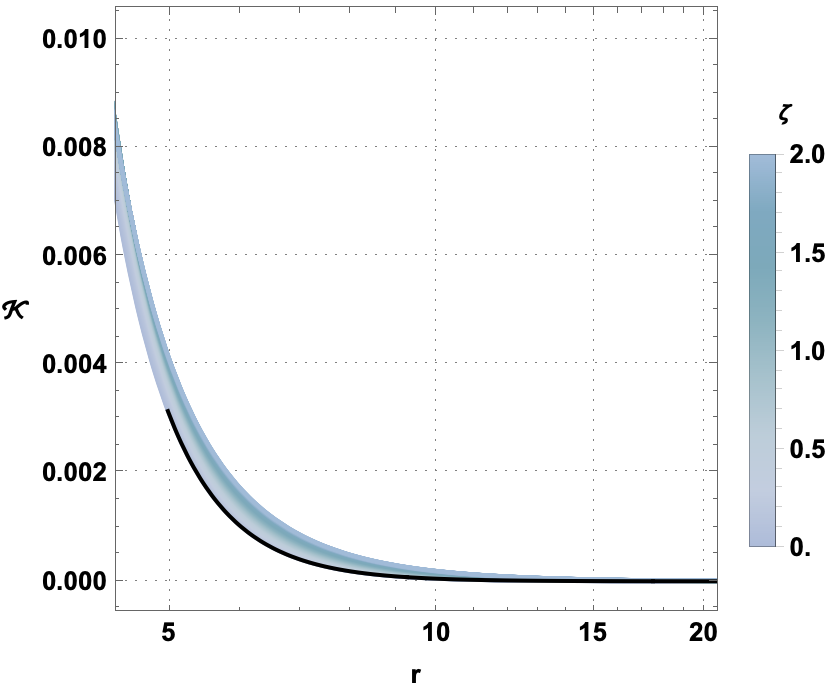}\>
\includegraphics[width=.44\textwidth]{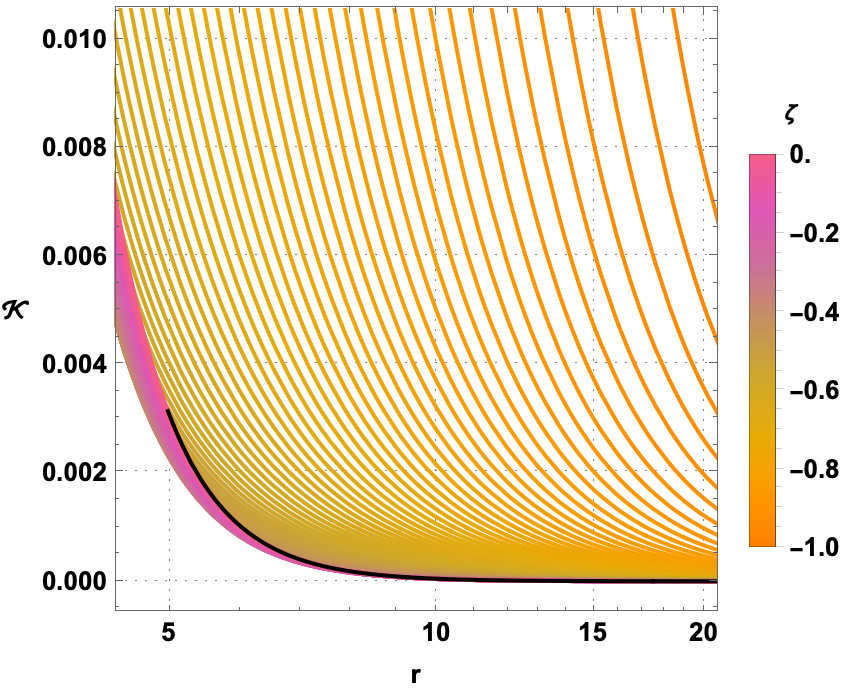}
\end{tabbing}
\caption{ \footnotesize   \it
Profile of the embedding diagram within different values of the correction parameter $\zeta$. {\bf Left: positive range} and {\bf Right: negative range}. The solid black line  corrreponds to  the Schwarzschild solution associated with   $\zeta=0$.
}
\label{fig2}
\end{figure}

Considering the event horizon (red circle), it decreases for negative values of $\zeta$, while for the positive ones, the event horizon increases to reach the value $\zeta=1.59$ then it  decreases.
 In addition, the Schwarzschild black hole is known to have a true physical singularity at $r=0$ and a coordinate singularity at $r=2M$. We now investigate whether the presence of  the Dunkl parameter affects the existence of these singularities or it can  play a role in regularizing the black hole at $r=0$. To do so,   a more comprehensive description of such  a black hole spacetime is needed. We recall the   Kretschmann invariant, which is a scalar quantity characterizing the curvature of spacetime at a specific point. It is defined as a combination of the Riemann tensor components
\begin{eqnarray}
    \mathcal{K}&=&R_{abcd}R^{abcd},\\ \nonumber
    &=&\frac{4 \left(\zeta ^2+2 (\zeta +1)^2 \left(2 \zeta  (\zeta +4)-\sqrt{8 \zeta +9}+9\right) M^2 r^{1-\sqrt{8 \zeta +9}}+4 \zeta  (\zeta +1) M r^{\frac{1}{2}-\frac{1}{2} \sqrt{8
   \zeta +9}}\right)}{(\zeta +1)^2 r^4}\,.
\end{eqnarray}
In Fig.\ref{fig2}, we depict the behavior of  such  a quantity in terms of the radial coordinate $r$  for  various ranges (positive/negative) of the Dunkl parameter $\zeta$.
%{\bf  new figures }

 It has been observed,  from the left panel  associated with the positive values of $\zeta$,  that the Kretschmann number increases as  $\zeta$ grows. This  indicates  that
 spacetime is significantly curved, leading to deviations from the flat spacetime  for the high values of the Dunkl parameter.  For negative ranges of $\zeta$, however,  two distinctive schemes appear. For $\zeta\leq-0.4$,  the Kretschmann quantity decreases as the absolute value of $\zeta$ increases, while for   $\zeta\geq-0.4$ the situation is inverted.

In the present work, we would like to go beyond such geometrical aspects by investigating the optical  behaviors  via the shadows, and the deflection angle computations  and other  related discussions.  This study could bring  data on the acceptable regions of the extended moduli space  by exploiting the known black  hole empirical  studies.  The forthcoming sections will concern the  geometric engineering of the   certain optical aspects   of the  Dunkl black hole via real curves and their contacts with  the EHT  empirical  findings.

 \section{ Optical behaviors of  the  Dunkl black hole}
 In this section, we conduct an examination of specific optical properties associated with the Dunkl black hole, focusing on certain  aspects such as  the shadows and the light deflection angle  behaviors in various contextual backgrounds.
\subsection{Shadows of  the Dunkl black hole}
\begin{figure}[!ht]
\centering 
			\begin{tabbing}
			\centering
			\hspace{10.5cm}\=\kill
\includegraphics[width=0.65\textwidth]{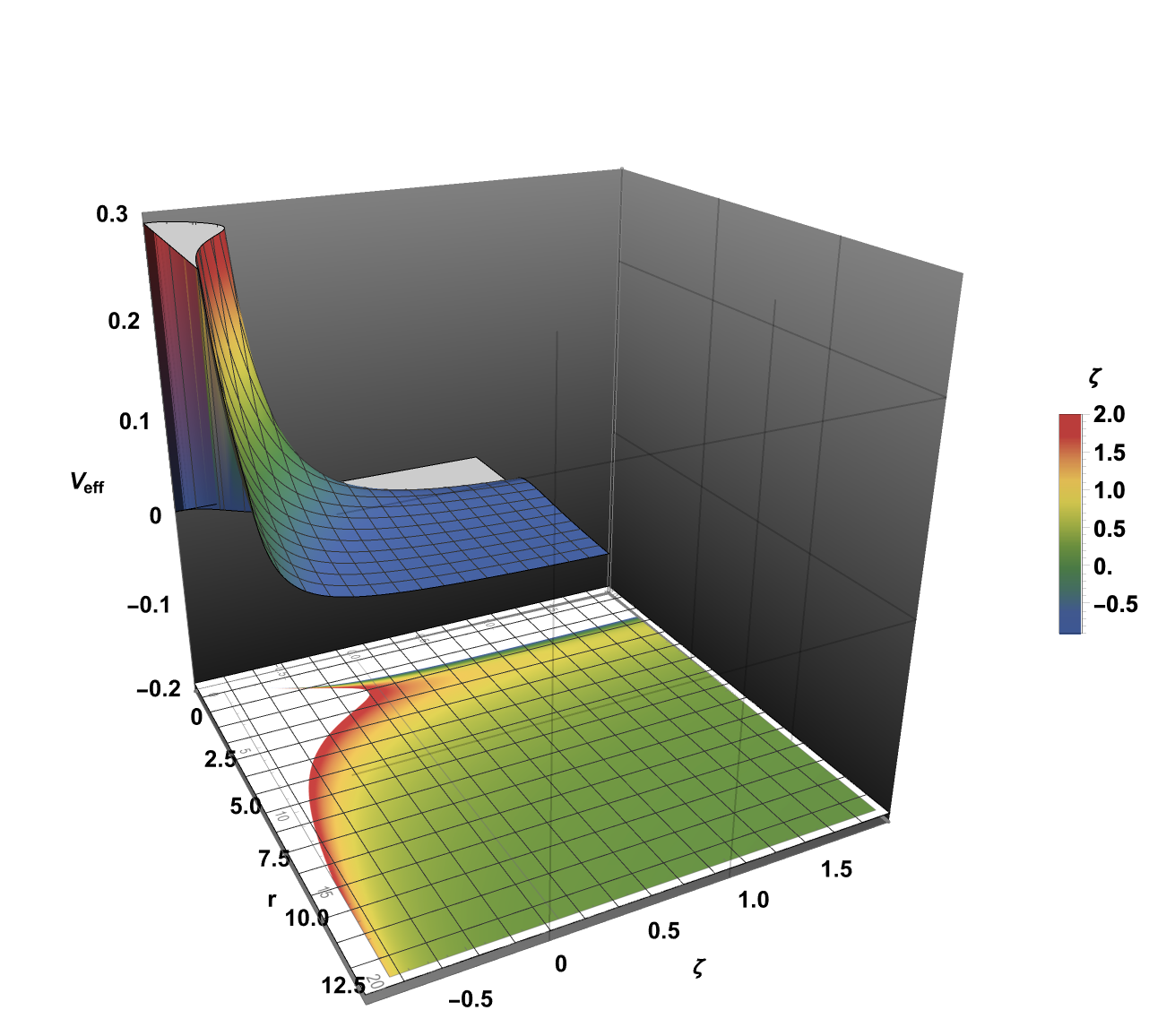}\>
\includegraphics[width=0.4\textwidth]{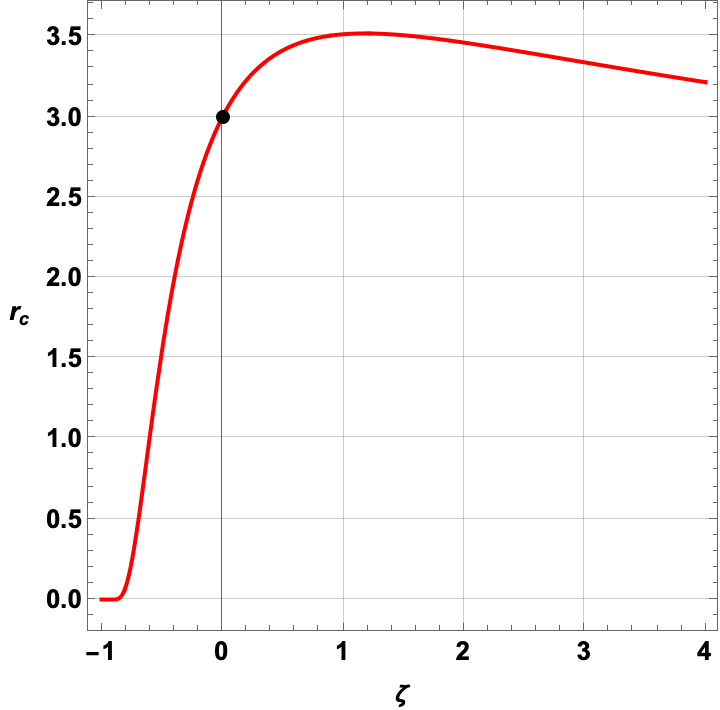}
% "\includegraphics" is very powerful; the graphicx package is already loaded
\end{tabbing}
\caption{\label{fig3x} \footnotesize   \it
{\bf Left: } Effective potential by varying the correction parameter $\zeta$. We have set $\mathcal{E}=0.01$ and $\mathcal{J}=1$. {\bf Right: } Critical radius $r_c$ in terms of the Dunkl parameter $\zeta$, with  $M=1$.
}
\label{pot}
\end{figure}

First,  we   would like to appraoch  the shadow  behavior  of   the Dunkl black hole. Precisely, we exploit  the Hamilton-Jacobi scenario needed to elaborate the equations of motion of photons near  such  a  black hole.  To start, we consider null geodesic equations 
\begin{equation}\label{hamij}
\mathcal{H}=\frac{1}{2}g^{\mu \nu}p_{\mu}p_{\nu}=0
\end{equation}
where the quantity  $g^{\mu \nu}$ represents the  associated  black hole metric given in terms of the line element $
ds^{2}= g^{\mu \nu}dx_{\mu}dx_{\nu}.
$ %For  simplicity reasons, we consider    a static spherically symmetric space-time background which can be described by the following  line element expression
%\begin{equation}
%\label{metricfunction}
%ds^{2}=f(r)dt^{2}-\frac{dr^{2}}{f(r)}-r^{2}d\theta^{2}-r^{2}\sin^{2}\theta d\phi^{2}
%\end{equation}
%where $f(r)$   denotes    the black hole function currying the essential data on the involved parameters.  
In the   equatorial plane defined by  $\theta=\pi/2$,  Eq.\eqref{hamij}  can be reduced to 
\begin{equation}\label{ramo}
\frac{1}{2}\left[-\frac{p_{t}^{2}}{f(r)}+f(r)p_{r}^{2}+\frac{p^{2}_{\phi}}{r^{2}}\right]=0.
\end{equation}
In these  computations, two constants of motion   appear naturally. They are  the energy $\mathcal{E}$  and  the angular momentum $\mathcal{J}$  of the photon   given by 
\begin{equation}
\mathcal{E}=-\frac{\partial \mathcal{H}}{\partial \dot{t}}\,, \qquad
\mathcal{J}=\frac{\partial \mathcal{H}}{\partial \dot{\phi}}.
\end{equation}
Here, one has used  the following equations
\begin{equation}\label{xx1}
\dot{t}=\frac{\partial H}{\partial p_{t}}=-\frac{p_{t}}{f(r)}\,,
\qquad
\dot{\phi}=\frac{\partial H}{\partial p_{\phi}}=\frac{p_{\phi}}{r^{2}}\,,
\end{equation}
where  the  over dot is the derivative with respect to  the affine parameter $\tau$. Using     the radial momentum $p_{r}$ via  the relation
\begin{equation}\label{xx2}
\dot{r}=\frac{\partial H}{\partial p_{r}}=p_{r}f(r)
\end{equation}
the   effective potential  satisfies 
\begin{equation}\label{ee1}
V_\text{eff}(r)+\dot{r}^{2}=0.
\end{equation}
This gives 
\begin{eqnarray}\label{e1}
 V_\text{eff}(r)=f(r)\left[\frac{\mathcal{J}^{2}}{r^{2}}-\frac{\mathcal{E}^{2}}{f(r)}\right]
 =\frac{\mathcal{J}^2 \left(\frac{1}{\zeta +1}-2 M r^{\frac{1}{2}-\frac{1}{2} \sqrt{8 \zeta
   +9}}\right)}{r^2}-\mathcal{E}^2,
\end{eqnarray}
where,  in the first line, $f(r)$  denotes the metric function given by Eq.(\ref{2}). 
Before going ahead, we illustrate the effective potential in the left panel of Fig.\ref{fig3x}.
It is straightforward to demonstrate that the effective potential for the Dunkel-Schwarzschild black hole reaches its maximum at the critical radius $r_c$
\begin{equation}
r_c=\left(\frac{\sqrt{8 \zeta +9}-3}{4\zeta  (\zeta +1)
   M}\right)^{-\frac{2}{\sqrt{8 \zeta +9}-1}}.
\end{equation}

\begin{figure}[!ht]
\centering
\hspace{2cm}
\begin{tikzpicture}
%\node[inner sep=0pt, text width=4cm ]  at (-4.5,0)
  %  {\includegraphics[width=2\textwidth]{Shadowfullzitan}};
\node[inner sep=0pt, text width=4cm ]  at (2,0)
    {\includegraphics[width=2.\textwidth]{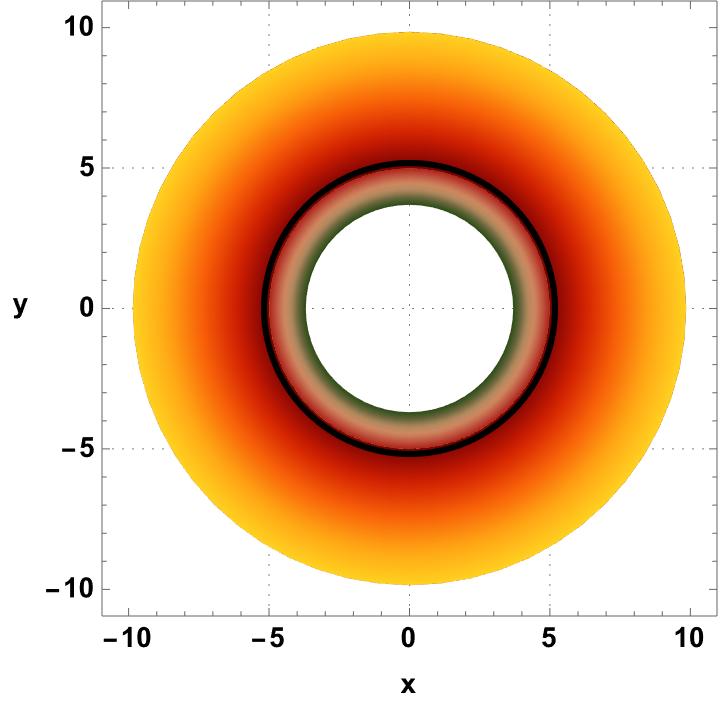}};

\node[inner sep=0pt, text width=4cm ]  at (10.5,0)
    {\includegraphics[width=1.2cm,height=6cm]{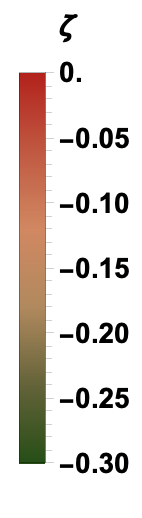}};

\node[inner sep=0pt, text width=4cm ]  at (12.,0)
    {\includegraphics[width=1cm,height=6cm]{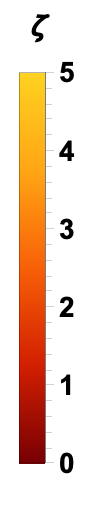}};
\end{tikzpicture}
        \caption{\footnotesize{\it Shadows  for  different values of the correction parameter $\zeta$.}}
        \label{figure1}
\end{figure}

For the Schwarzschild solution,  the effective potential $V_{eff}$ for a photon reaches its maximum at $r=3M$, indicating an unstable circular orbit,  as $r$ approaches infinity. The effective potential asymptotes to a constant value. As shown in both panels of Fig.\ref{fig3x}, as the parameter $\zeta$ increases, the critical radius $r_c$ initially rises, reaching a peak around $\zeta=1.18$, before it starts to decrease. 
This behavior suggests that as the Dunkl parameter grows, the size of the unstable circular orbits initially expands, then  it contracts. The photon orbits which   are associated with the maximum effective potential are circular and unstable. These unstable circular orbits define the boundary of the black hole apparent shape and can be determined by maximizing the effective potential required by 

%To determine the  black hole shadow illustrations, one needs to identify the radius of the photon sphere. This can be determined    by use of  the   following  constraints
\begin{equation}\label{effpho}
V_\text{eff}(r_c)=0,\quad  \text{ and } \left.\frac{\partial V_\text{eff}(r)}{\partial r}\right|_{r=r_c}=0.%\quad \text{and}\quad  \left.\frac{\partial^2 V_{e}(r)}{\partial r^2}\right|_{r=r_p}>0.
\end{equation}
The motion of photons in the  Dunkl spacetime is governed by Eqs.\eqref{xx1}-\eqref{ee1}. The behavior of photons near the black hole is characterized by two impact parameters, depending  on the constants of motion.  For general orbits around the black hole, these impact parameters are given by
%XXXXXX
%Using  Eq.\eqref{ramo},  one obtains the equation associated  with  the photon orbital 
%\begin{equation}
%\frac{dr}{d\phi}=\pm r\sqrt{f(r)\left[\frac{r^{2}\mathcal{E}^{2}}{f(r)\mathcal{J}^{2}}-1\right]}.
%\end{equation}
%The turning point of the photon orbit constrained  by  $\left.\frac{dr}{d\phi}\right|_{r=R}=0$  leads to 
%\begin{equation}\label{eosf}
%\frac{dr}{d\phi}=\pm r\sqrt{f(r)\left[\frac{r^{2}f(R)}{f(r)R^{2}}-1\right]}.
%\end{equation} 
\begin{equation}
\xi=\frac{L}{E}, \qquad \text{ and }\qquad\eta =\frac{\kappa}{E^{2}},
\end{equation} 
where $ \kappa$ is a separating constant where $L$ and $E$ are constants of motion.  They  are needed to   visualize  the  shadow geometries  in  four dimensions. To obtain  an elegant representation, however,  one could use    the celestial coordinates  $(X,Y)$   representing   all projections of  the spherical photon orbits \cite{Vazquez:2003zm,Eiroa:2017uuq}.    They read as 
\begin{eqnarray}
X & =& -r_o\frac{p^{(\phi)}}{p^{(t)}},\\
Y & =& r_o\frac{p^{(\theta)}}{p^{(t)}}.
\end{eqnarray}
 where $r_{o}$   denotes   the distance of the observer from the black hole\cite{Cunha:2016bpi,Zhang:2020xub}.  Precisely,  they are given by 
\begin{eqnarray}
\small
X&=& \lim_{r_{_{o}\rightarrow +\infty}} \left( -r^2_{o} \sin \theta_{o} \frac{d\phi}{d r}\right), \\
Y &=&  \lim_{r_{_{o}\rightarrow +\infty}}\left( r^2_{o}\frac{d\theta}{dr}\right) 
\end{eqnarray}
where  $\theta_{o}$  is   the angle of  the inclination between the observer line of  sight and the axis of the black hole rotation.  Instead of giving an analytic discussion, we use  a  numerical   one by varying the Dunkl parameter  $\zeta$. %and $\Lambda$.
%First, we  examine  the  impact of the cosmological constant on the shadows. In Fig.\ref{figure2},  we illustrate the  shadows with negative and positive cosmological constant
%values. 
%\begin{figure}[!ht]
%\begin{minipage}{.6\linewidth}
%\centering
%%\subfloat[Contours illustrating the dependence of the angular diameter of the shadow of M87*]{\label{fig2.1}
%\includegraphics[scale=0.8]{Shadowfulllambda}
%\end{minipage}%
%\begin{minipage}{.2\linewidth}
%\centering
%%\subfloat[Shadow radius for dunkel black hole with Sgr $A^*$ observed datas ]{\label{fig2.2}
%\includegraphics[scale=1.4]{BLlambda}
%\end{minipage}\par\medskip
%        \caption{\footnotesize{\it The profile of the shadow within different values of  cosmological constant $\Lambda$.}}
%        \label{figure2}
%\end{figure}
%Since we are considering the non rotating solutions,   we  obtain   circular shadow geometries. 
%It has been observed from 
%this figure that the shadow sizes increase with  $\Lambda$. As expected,  the shadow size is larger for the  geometry with positive cosmological constant. This result matches perfectly with the previous one \cite{}.
%Now, we move to discuss the relevant parameter in the present work.  
 Fig.(\ref{figure1})   presents  the shadows   as a  function of  such a relevant parameter  $\zeta$ in the present work.

It follows from this figure    that  $\xi$    can be interpreted as a geometric deformation  parameter  controlling   the involved sizes.  Precisely, the size     increases by increasing  $\xi$ being  a relevant  free parameter.   However, the empirical findings could be exploited to  impose constraints  on such a parameter needed to predict certain  accepted physical   ranges.   
%\newpage
Further, one  utilizes the numerical backward ray-tracing approach  to analyze the black hole shadows \cite{Hu:2020usx}. This technique provides a visual representation of the shadow extent within the lensing ring. In the following tetrad
\begin{equation}\label{zjbj}
\begin{aligned}
e_{(t)}=&\left.\frac{1}{\sqrt{-g_{tt}}}\pp_t\right|_{(r_0,\theta_0)},
e_{(r)}=&\left.-\frac{1}{\sqrt{g_{rr}}}\pp_r\right|_{(r_0,\theta_0)},
e_{(\theta)}=&\left.\frac{1}{\sqrt{g_{\theta\theta}}}\pp_\theta\right|_{(r_0,\theta_0)},
e_{(\varphi)}=&-\left.\frac{1}{\sqrt{g_{\phi\phi}}}\pp_\phi\right|_{(r_0,\theta_0)},
\end{aligned}
\end{equation}
$e_t$ represents the observer  four-velocity, $e_\varphi$ is directed toward the black hole center.  However,  $e_t \pm e_\varphi$ squares with  the principal null directions of the metric \cite{Zhang:2020xub,Chakhchi:2024tzo}. For each light ray parameterized as $ \lambda(s) $, with coordinates given by $ t(s), r(s), \vartheta(s), \varphi(s) $, the tangent vector takes the general form
\begin{align}\label{lambda1}
\dot{\lambda}=\dot{t}\pp_t+\dot{r}\pp_r+\dot{\vartheta}\pp_\vartheta+\dot{\varphi}\pp_\phi.
\end{align}
Using an  observer frame, the tangent vector of the null geodesic is  found  to be \begin{equation}
\dot{\lambda}=|\overrightarrow{OP}|(- \chi e_0+\sin\theta\cos\psi e_1+\sin\theta\sin\psi e_2+\cos\theta e_3),
\end{equation}
where  $\chi$ is a scale  factor. In a  three-dimensional space,  the  vector  $|\overrightarrow{OP}|$ represents the tangent vector of the null geodesic at    $O$.  The stereographic representation of the celestial sphere on a plane is 
then re-established
\begin{equation}\label{B4}
\begin{aligned}
x_{P'}=&-2|\overrightarrow{OP}|\tan\frac{\theta}{2}\sin\psi, \ 
y_{P'}=&-2|\overrightarrow{OP}|\tan\frac{\theta}{2}\cos\psi,
\end{aligned}
\end{equation}
where  $P'$ denotes the projection of  $P$ onto the Cartesian plane.
  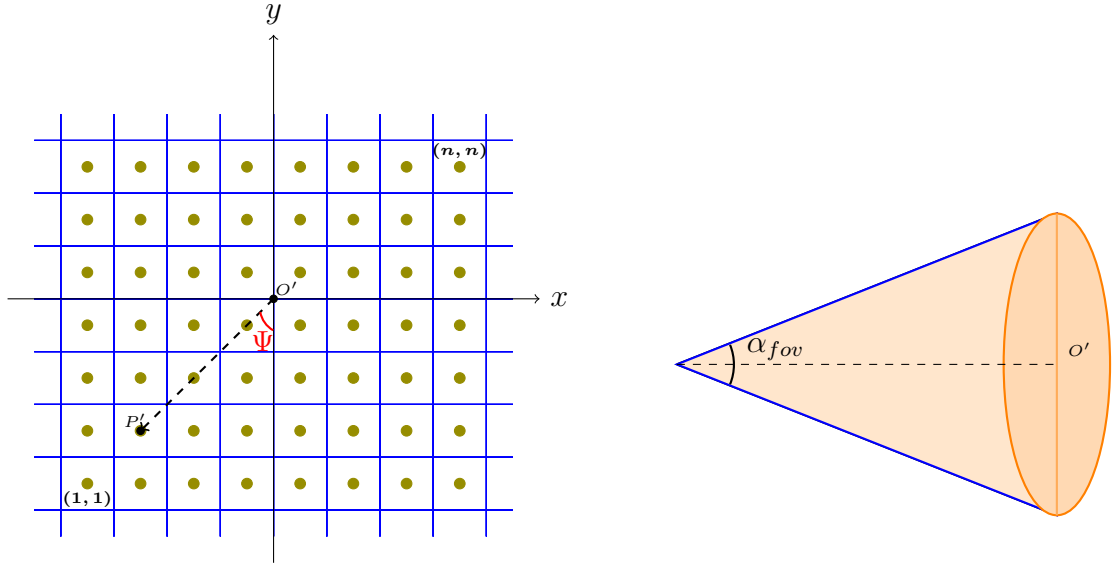
\begin{figure}[!ht]
\centering 
			\begin{tabbing}
			\centering
			\hspace{9.cm}\=\kill
\begin{tikzpicture}[scale=.7]
    % Define grid dimensions
    \def\gridWidth{9}  % Number of squares along the x-axis
    \def\gridHeight{8} % Number of squares along the y-axis
    \def\squareSize{1} % Size of each square

    % Draw grid with smaller green circles in each square, excluding the right and top borders
    \foreach \x in {-4,...,4} {
        \foreach \y in {-4,...,3} {
            % Draw grid lines in orange
            \draw[blue, thin] (\x*\squareSize,-4.5) -- (\x*\squareSize,3.5); % Vertical lines
            \draw[blue, thin] (-4.5,\y*\squareSize) -- (4.5,\y*\squareSize); % Horizontal lines
            
            % Place smaller green circle in each square, excluding right and top borders
            \ifnum\x<4
                \ifnum\y<3
                    \filldraw[olive] (\x*\squareSize + 0.5*\squareSize, \y*\squareSize + 0.5*\squareSize) circle (0.1);
                \fi
            \fi
        }
    }

    % Define coordinates for labeled points
    \coordinate (O') at (0,0);  % Origin
    \coordinate (P') at (-2.5,-2.5);  % Point P' (left bottom)

    % Draw centered axes
    \draw[->] (-5,0) -- (5,0) node[right] {$x$};
    \draw[->] (0,-5) -- (0,5) node[above] {$y$};

    % Mark and label main points
    \filldraw (O') circle (2pt) node[below left] {};
    \filldraw (P') circle (2pt) node[below left] {};

    % Draw angle arc between the y-axis and O'P' to indicate the angle Psi
    \draw[->,dashed,thick] (O') -- (P') node[midway, below right,red] {};
    \draw[thick,red] (-.25,-.25) arc[start angle=190,end angle=240,radius=0.5];

    % Label the arc indicating position of pixels    
    \node at (-3.5,-3.8) {\tiny $\bm{(1,1)}$};
    \node at (3.5,2.8) {\tiny$\bm{(n,n)}$};
    \node at (-2.6,-2.3) {\tiny$P'$};
     \node at (0.25,0.2) {\tiny$O'$};
      \node[red] at (-0.2,-.8) {\small$\Psi$};
    
\end{tikzpicture}
\hspace{1cm}
 \begin{tikzpicture}

    % Define the apex of the cone
    \coordinate (A) at (0,0);

    % Define points for the base of the cone
    \coordinate (B) at (5,2);
    \coordinate (C) at (5,-2);

    % Draw the cone edges
    \draw[thick] (A) -- (B);
    \draw[thick] (A) -- (C);

\draw[thick,blue,fill=orange!20] (A) -- (B) -- (C) -- cycle;
% Draw the circular base of the cone as a full ellipse using pgfellipse
    \fill[orange!30] (5,0) ++(0,0) [rotate=0] ellipse [x radius=0.7, y radius=2];
    \draw[thick, orange] (5,0) ++(0,0) [rotate=0] ellipse [x radius=0.7, y radius=2];

    % Draw the dashed axis of symmetry
    \draw[dashed] (A) -- (5.,0);
     \draw[thick,orange!65] (5,2) -- (5.,-2);

    % Draw the arc representing the angle theta
    \draw[thick] (0.7,-0.265) arc[start angle=-22, end angle=22, radius=0.7];
    \node at (1.3,0.2) {$\alpha_{fov}$};
    \node at (5.3,0.2) {\tiny$O'$};

    % Add the solid angle formula
   \node[purple] at (2.5,-2.5) {$  $};

\end{tikzpicture}
\end{tabbing}
\caption{\footnotesize\it {\bf Left :}Illustration of the pixels, where a classical Cartesian coordinate with origin $ O' $ is used. {\bf Right :} Illustration of the field of view.}\label{pixel}
   \end{figure} 
 We establish a standard Cartesian coordinate system centered at the origin $O',$ as illustrated in the left panel of Fig.\ref{pixel}. We then define the  view  field  denoted by angles within the $yO'z$ and $xO'z$ planes, which we assume to be equal for simplicity. 
 The right panel of Fig. \ref{pixel}  provides an illustrating  example of  the associated geometric  configurations. The length $L$ of the square projection screen is then calculated as follows
 \begin{align}
L=2|\overrightarrow{OP}|\tan\frac{\alpha_\text{~fov}}{2}.
\end{align} 
We define a screen with $n\times n$ pixels, each occupying a length given by
  \begin{align}
\ell=\frac{2|\overrightarrow{OP}|}{n}\tan\frac{\alpha_\text{~fov}}{2}.
\end{align}
Pixels are indexed by $(i,j)$, where $(1,1)$ corresponds to the bottom left corner and $(n,n)$ to the top right corner,  where  $i$ and $j$  take values  from $1$ to $n$. The Cartesian coordinates for the center of each pixel are given by
 \begin{align}\label{B7}
x_{P'}=\ell \left(i-\frac{n+1}{2}\right), && y_{P'}=\ell\left(j-\frac{n+1}{2}\right).
\end{align}  
 The pixels  $ (i,j) $ are related to the angles $ (\theta,\Psi) $ as follows
\begin{equation}
\begin{aligned}
\tan\Psi&=\frac{j-(n+1)/2}{i-(n+1)/2},\ 
\tan\frac{\theta}{2}&=\tan\frac{\alpha_\text{~fov}}{2}\frac{\sqrt{[i-(n+1)/2]^2+[j-(n+1)/2]^2}}{n}.
\end{aligned}
\end{equation}
In a cross-sectional view of a "large sphere" that encloses both the black hole and the observer, revealing its internal configuration as illustrated in  Fig.\ref{lensing}. The sphere is mapped with a grid of latitude and longitude lines spaced at intervals of $\pi/18$. To display the image of the black hole, a specific color is assigned to a segment of the grid, corresponding to one of four sections of the extended source. The color of each pixel in the resulting image is is then determined by plotting photon trajectories from the associated source points.  Dark regions in the image signify photon absorptions by the black hole.

\begin{figure}[!ht]
\centering 
			\begin{tabbing}
			\centering
			\hspace{10.5cm}\=\kill
\includegraphics[width=0.2\textwidth]{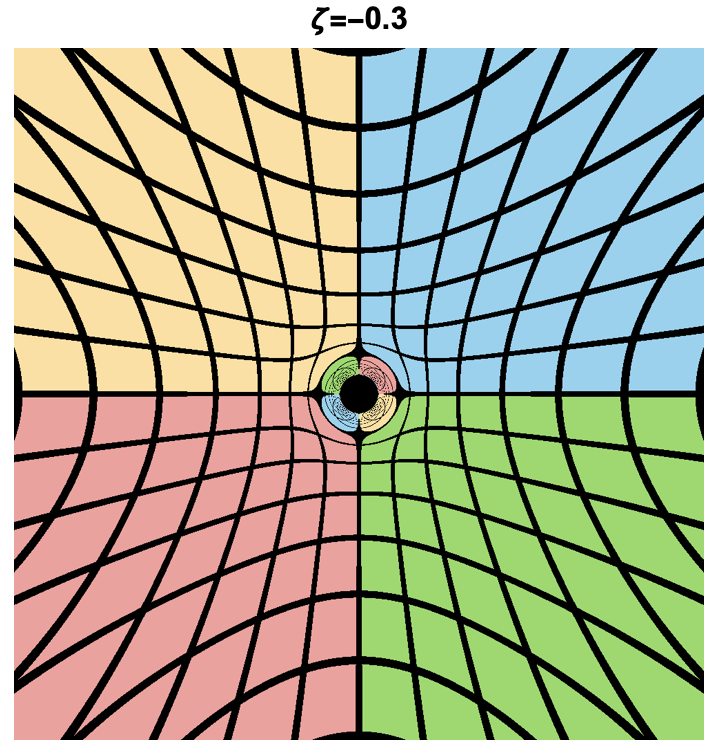}
\includegraphics[width=0.2\textwidth]{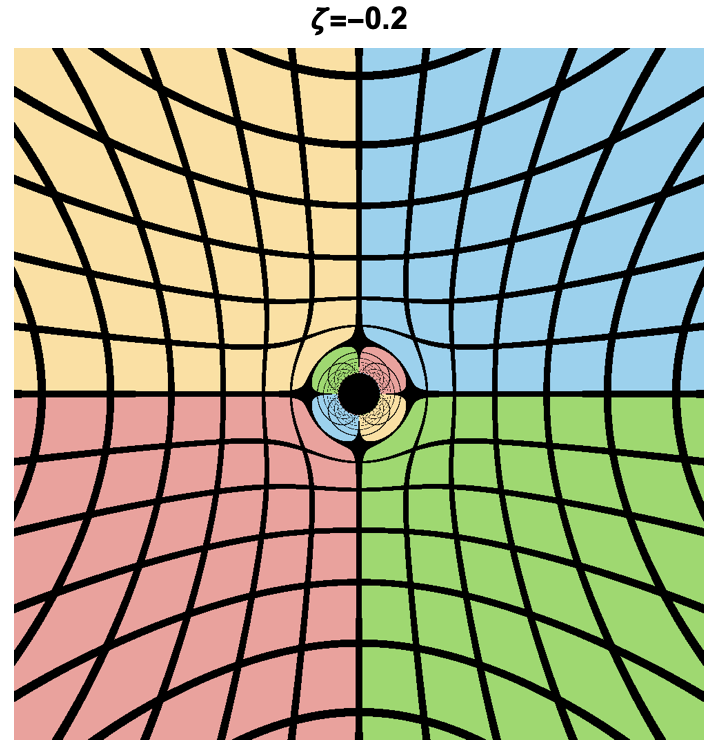}
\includegraphics[width=0.2\textwidth]{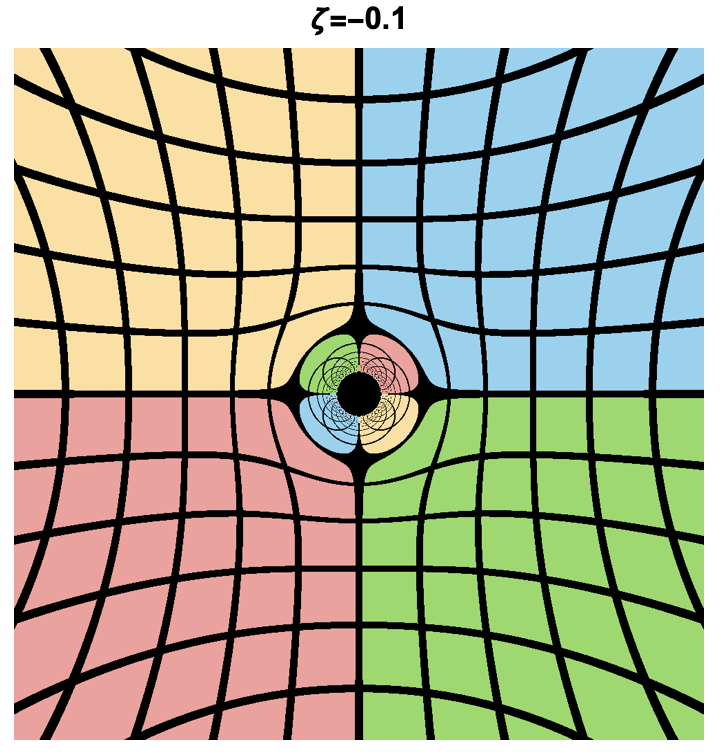}
\includegraphics[width=0.2\textwidth]{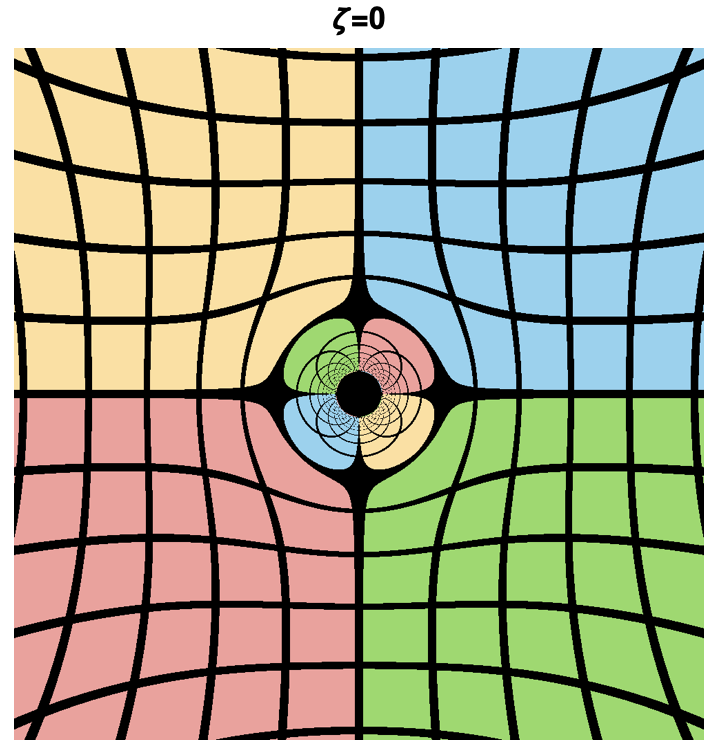}
\includegraphics[width=0.2\textwidth]{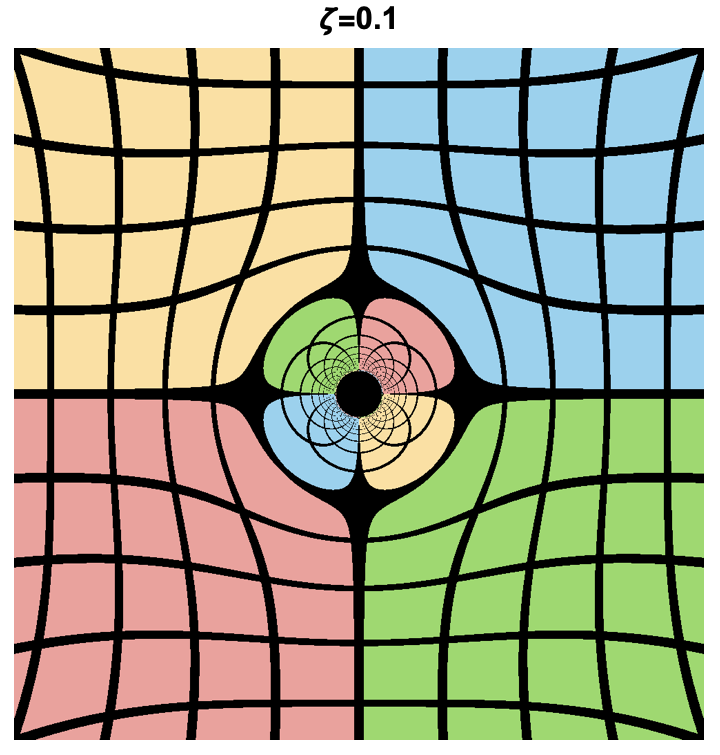}\\
\includegraphics[width=0.2\textwidth]{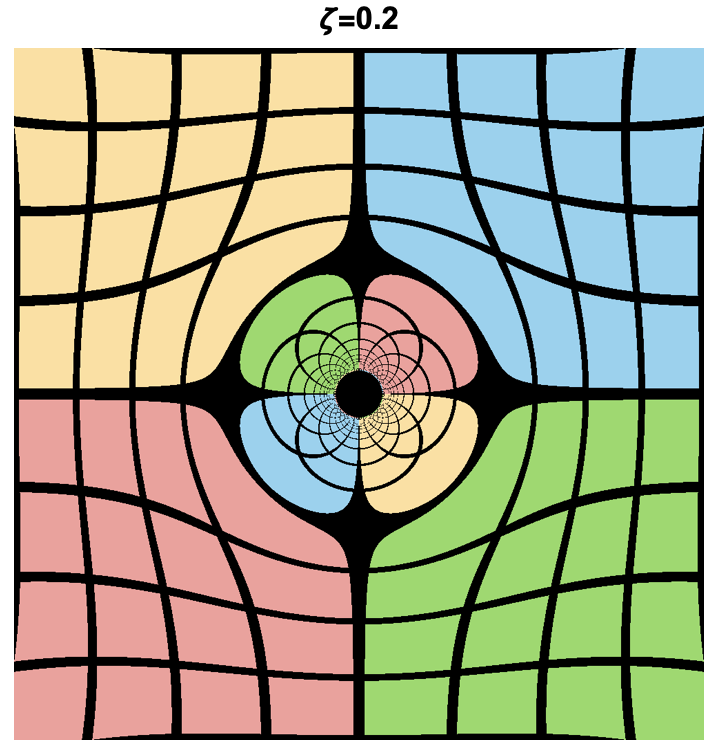}
\includegraphics[width=0.2\textwidth]{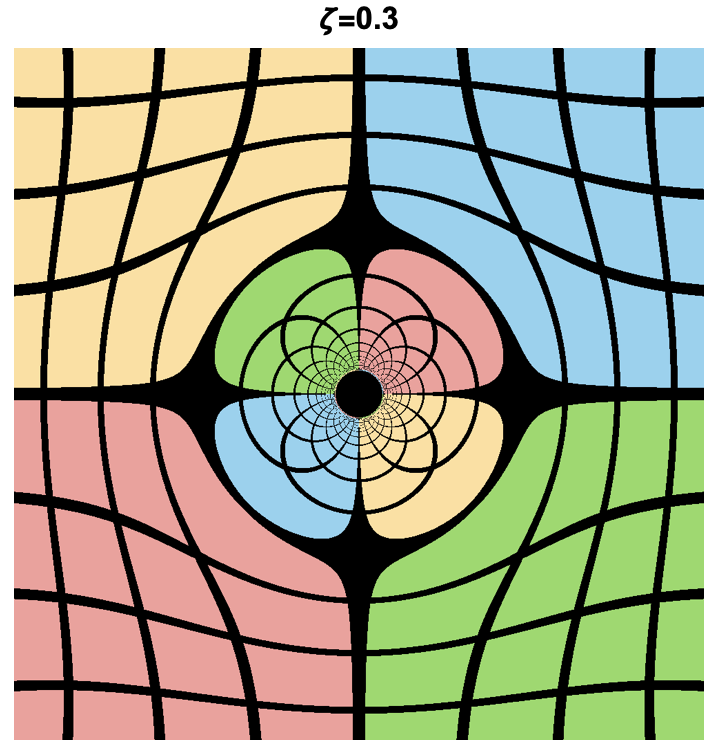}
\includegraphics[width=0.2\textwidth]{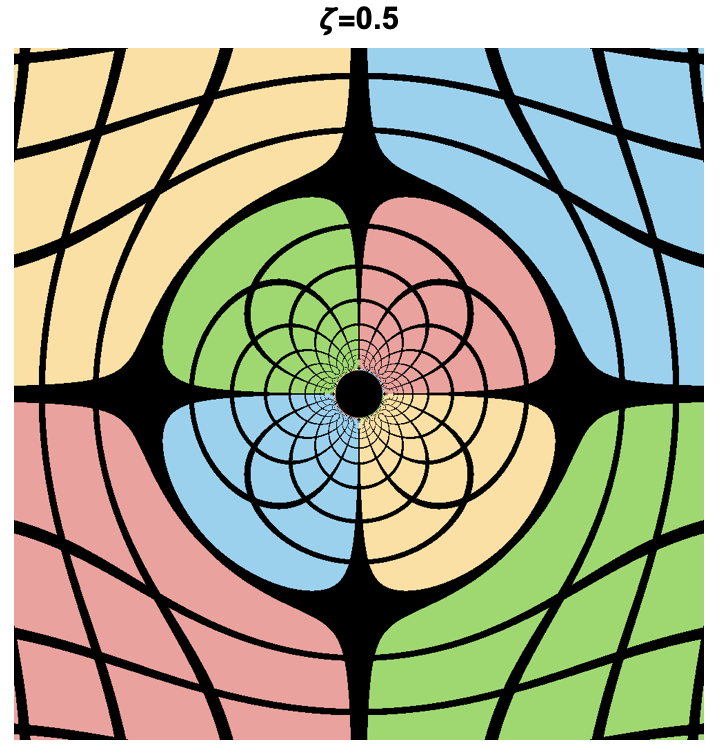}
\includegraphics[width=0.2\textwidth]{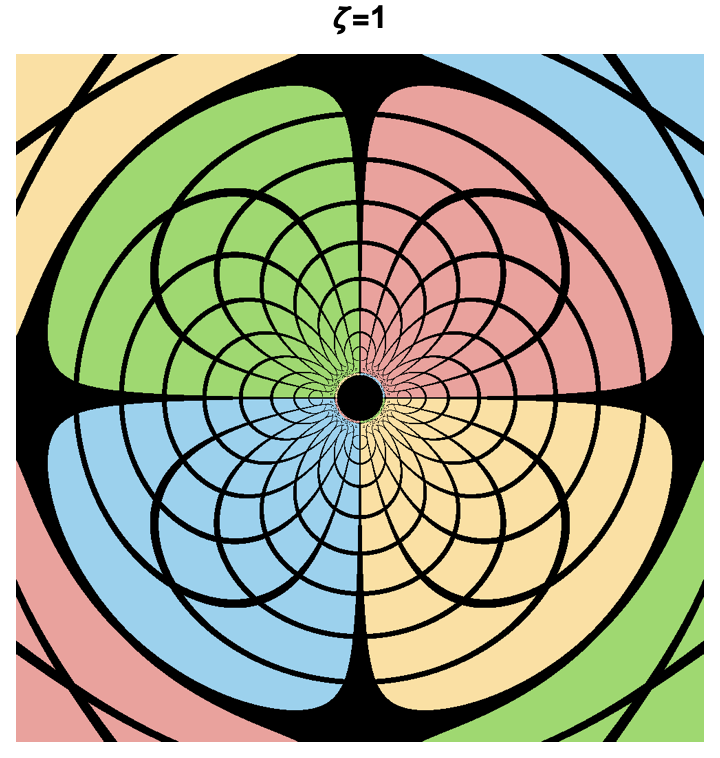}
\includegraphics[width=0.2\textwidth]{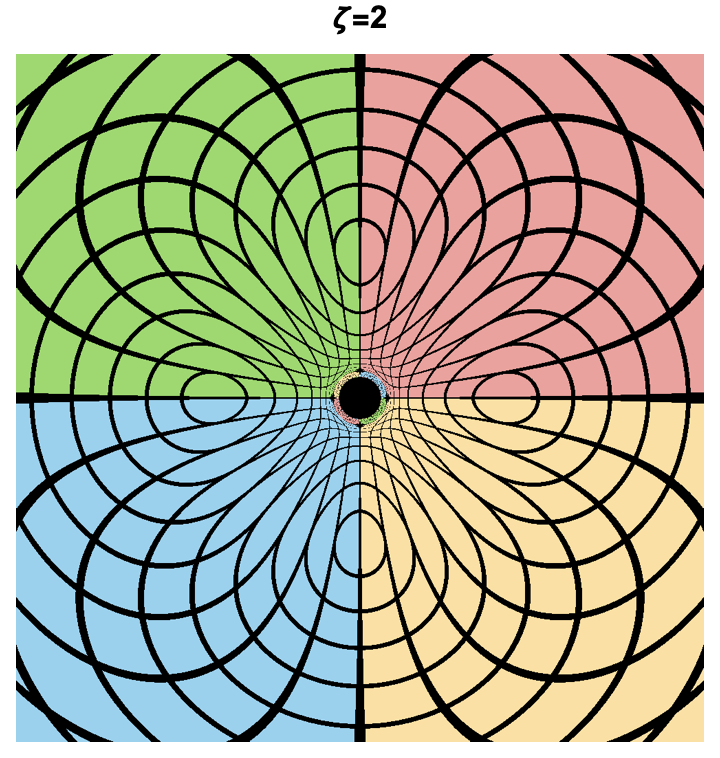}

% "\includegraphics" is very powerful; the graphicx package is already loaded
\end{tabbing}
\caption{\label{fig3} \footnotesize   \it
Shadows and lensing rings of the Dunkl black hole by taking the  various  $\zeta$ values, as seen by an observer at $\theta = \pi/2$.}
\label{lensing}
\end{figure}
Fig.\ref{lensing} reveals a notable geometric deformation when the Dunkl parameter $\xi$ deviates significantly from zero, while the shadow radius aligns consistently with the behavior observed in Fig.\ref{figure1}.

%\begin{figure}[!ht]
%\centering % \begin{center}/\end{center} takes some additional vertical space
%\includegraphics[width=0.7\textwidth]{Potential}
% "\includegraphics" is very powerful; the graphicx package is already loaded
%\caption{\label{fig1} \footnotesize   \it
%The profile of the effective potential within different values of the correction parameter $\zeta$.
%}
%\end{figure}

%The Lyapunov exponent of the
%unstable null circular geodesic is given by \cite{PhysRevD.79.064016}
%\begin{equation}
%  \lambda=\sqrt{\frac{r_{\text{o}}^{2}f\left(  r_{\text{o}}\right)  }{L^{2}%
%  }V_{\text{eff}}^{\prime\prime}\left(  r_{\text{o}}\right)  }, \label{lambda}%
%\end{equation}

%\begin{figure}[!ht]
%\centering % \begin{center}/\end{center} takes some additional vertical space
%\includegraphics[width=0.7\textwidth]{lyapunov}
% "\includegraphics" is very powerful; the graphicx package is already loaded
%\caption{\label{fig1} \footnotesize   \it
%The profile of the lyapunov exponent within different values of the correction parameter $\zeta$.
%}
%\end{figure}
\newpage

 To furnish  the accretion disk images,  one should consider  a screen with sides of $25M$. This screen,  being positioned at $ r_o=100M$  with  asymptotic infinity  behaviors for practical purposes, is oriented perpendicular to the observer line of sight toward the black hole. It is split into $500\times500$ pixels. In order to  scrutinize  different accretion disk configurations,   the  observer inclination angles  are considered to  $\iota=65^\circ$ and $85^\circ$.     Null geodesics are created by working backwards in time from each pixel on the screen to produce the images.

Tab.\ref{figure8}  exposes  the accretion disk images produced  by considering different values the Dunkl parameter $\zeta$ and the  observer inclination angle $\iota$.

\begin{table}[!ht]
        \centering
        \begin{tabular}{p{0.5cm}p{3cm}p{3cm}p{3cm}p{3cm}p{3cm}}
           \toprule
           \centering
            $\iota$/$\zeta$. & $\zeta=-0.1$ & $\zeta=-0.05$ &  $\zeta=0$ & $\zeta=0.2$ & $\zeta=0.5$\\
            \midrule
            $85^\circ$ & \includegraphics[scale=.27]{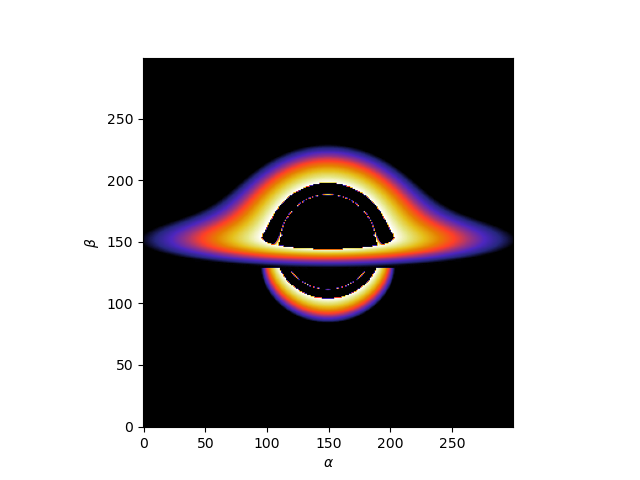} & \includegraphics[scale=.27]{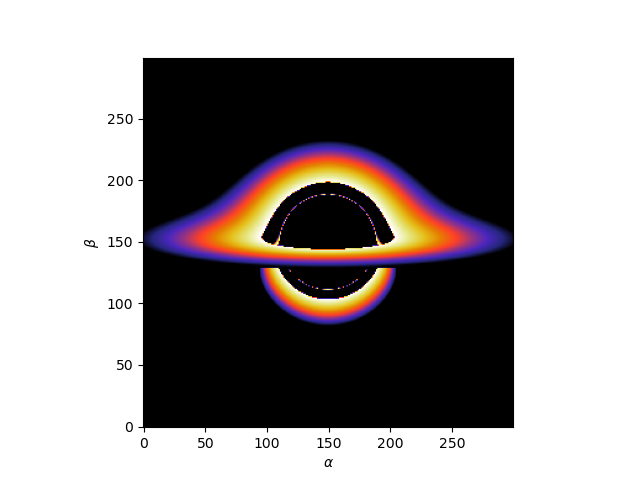} & \includegraphics[scale=.27]{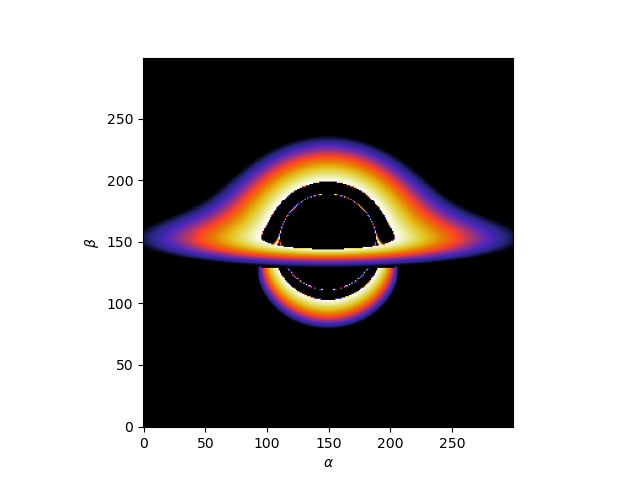} & \includegraphics[scale=.27]{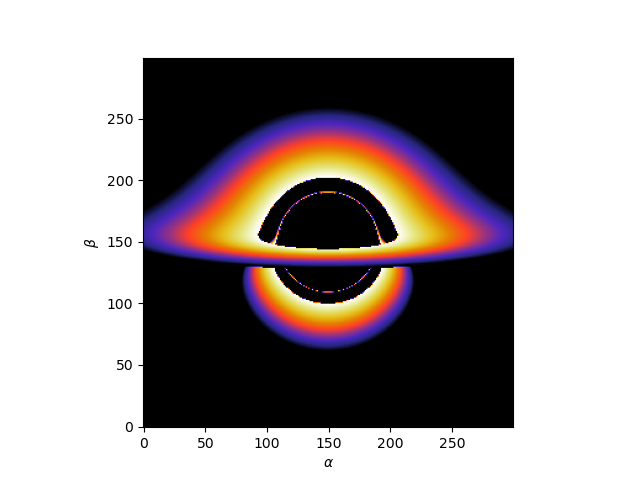} &
            \includegraphics[scale=.27]{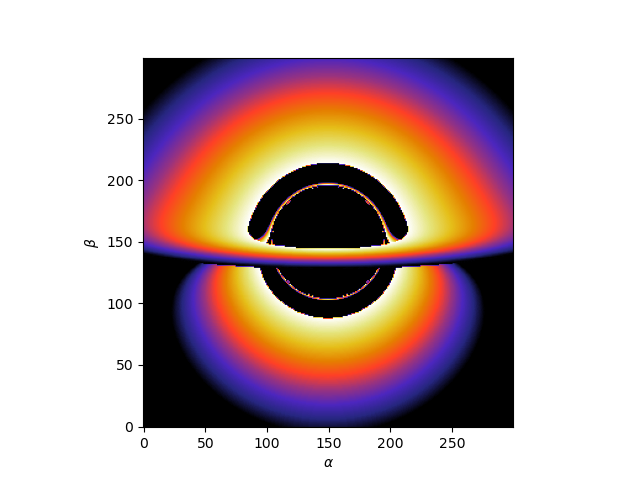}\\
            $60^\circ$ & \includegraphics[scale=.27]{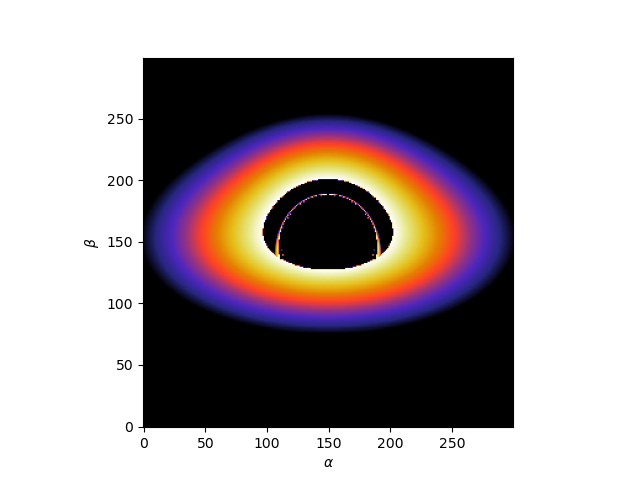} & \includegraphics[scale=.27]{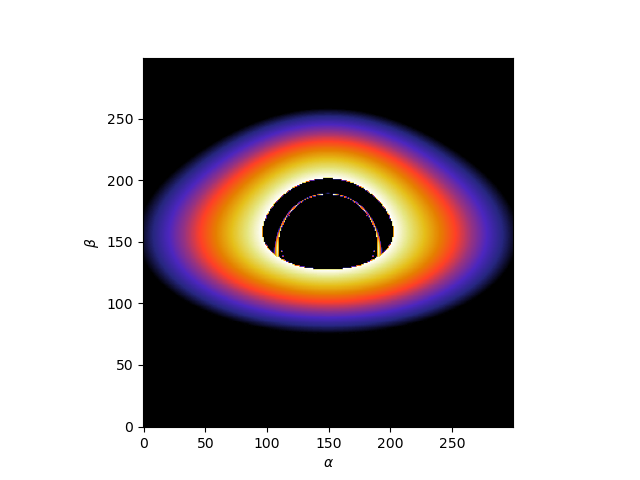} & \includegraphics[scale=.27]{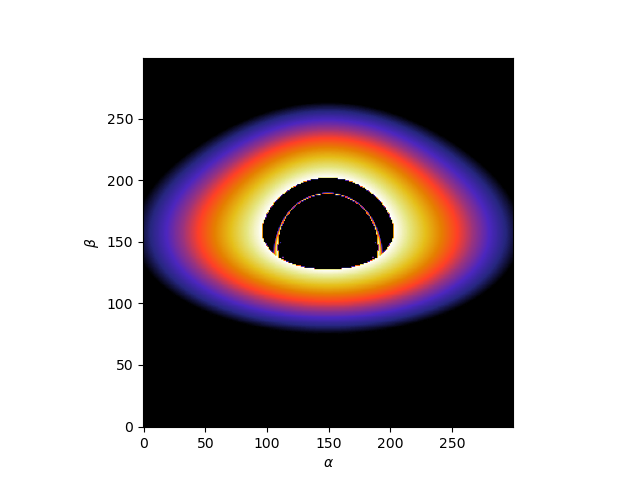} & \includegraphics[scale=.27]{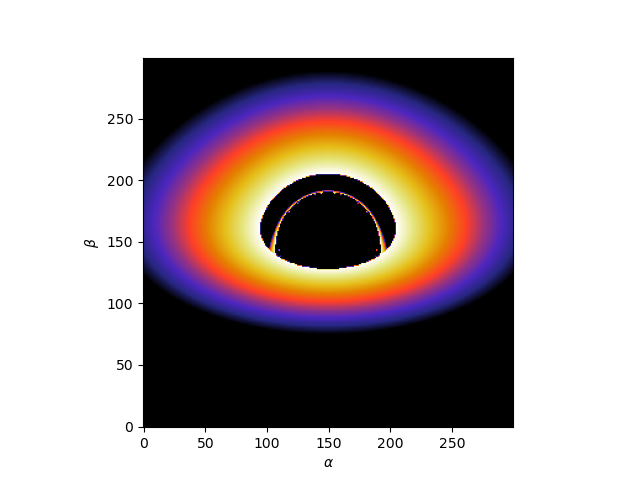} & \includegraphics[scale=.27]{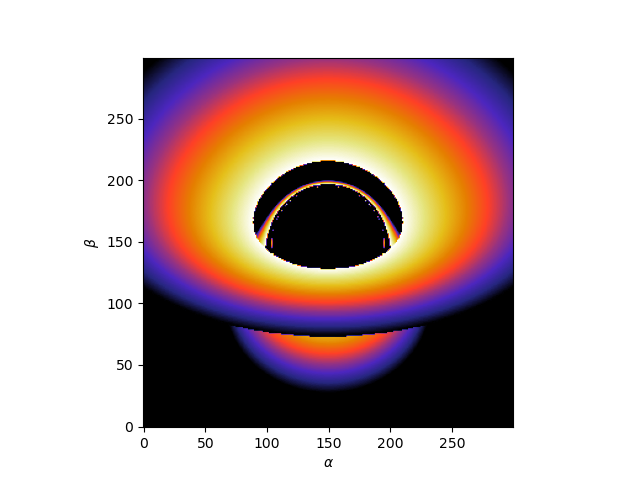} \\
            
            \bottomrule
        \end{tabular}
        \caption{\footnotesize{\it   Shadows cast by photons tracing null geodesics of the Dunkl black hole by varying  $\zeta$.}}
        \label{figure8}
    \end{table}

\newpage
 The morphology of the resulting images of the Dunkl black hole closely looks like those of  the Schwarzschild one, although the paramater $\zeta$  significantly influences their appearances.

\subsection{Lyapunov exponent} 
In this part, we deal with   the Lyapunov exponent behaviors. It is recalled that  the Lyapunov exponents evaluate how the close trajectories in  the phase space can be  developed. This can  quantify  the average rate of their divergence or convergence. A positive exponent indicates the  divergence behavior, reflecting a high sensitivity to initial conditions. Moreover,  the geodesic stability analysis is evaluated throughout the terms of  the Lyapunov exponents. Therefore, the  unstable circular orbits are intricately connected to the  chaotic dynamics around the  black holes, making their study valuable for understanding the chaos  nature. For instance, Maldacena, Shenker, and Stanford proposed a conjecture suggesting a universal upper limit for Lyapunov exponents $\lambda$, given by 
\begin{equation}
\lambda\leq \frac{2\pi T}{\hbar},
\end{equation}
where $T$ represents the system temperature \cite{Maldacena:2015waa}. However, as demonstrated in \cite{Lei:2021koj}, this bound can be exceeded in the case of unstable circular displacements of charged particles in the vicinity of a charged black hole. Furthermore,  the unstable circular null geodesics form photon spheres surrounding the event horizon, which are crucial for black hole observational studies.

\begin{figure}[!ht]
\centering 
\includegraphics[width=.6\textwidth]{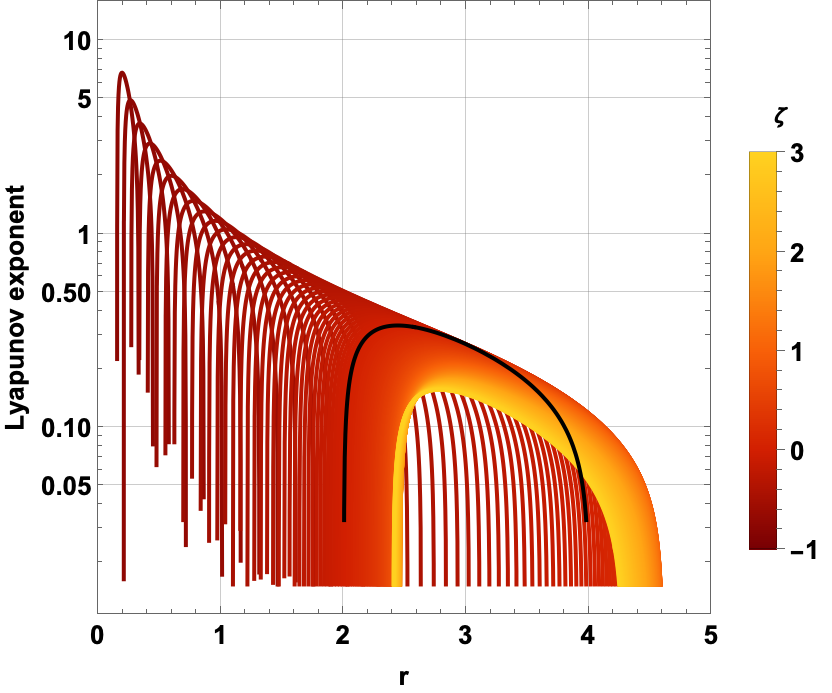}
\caption{\label{fig1} \footnotesize   \it
Logarithmic profile of the Lyapunov exponent within different values of the Dunkl parameter $\zeta$, with  $M=1$.
}
\label{lyapunov}
\end{figure}

According to \cite{Cardoso:2008bp},  the Lyapunov exponents of the null circular geodesic with unstable motion is expressed in terms of the effective potential  given in Eq.\eqref{e1}  as  follows 
\begin{equation}
  \lambda=\sqrt{\frac{r_{\text{c}}^{2}f\left(  r_{\text{c}}\right)  }{L^{2}%
  }V_{\text{eff}}^{\prime\prime}\left(  r_{\text{c}}\right)  }. \label{lambda}%
\end{equation}
In Fig.\ref{lyapunov}, we illustrate  the logarithmic variation of the Lyapunov exponent in terms of $r$ within various values of  the Dunkl parameter $\zeta$.
The present analysis indicates that the maximum instability tends to rise when there is a reduced contribution from the Dunkl parameter. Conversely, as the variable $r$ increases, the level of instability generally decreases. It is important to highlight that these findings are highly sensitive to the underlying geometric framework. This sensitivity suggests that  the present  investigation   may serve as a significant tool in differentiating between the  Schwarzschild black hole solution   and  the Dunkl black hole.
\subsection{Spherically infalling accretion}

It has been suggested  that an  accretion disk arises from a gravitational force  of a central object on the surrounding matter  such as gas, dust, plasma, including dark matter\cite{Jaroszynski:1997bw,Bambi:2012tg,Bambi:2013nla,Gralla:2019xty}. As this matter converges towards the center, it gains a kinetic energy producing a promptly  rotating disk. The  associated disk  temperature and the  density  emit  certain  radiation forms, such as X-rays, visible lights, or radio waves. These disks play a primordial  role in facilitating mass and angular momentum transfer during the gravitating object evaluations.
When an accretion disk is aligned with the gravitational lensing effect, it  generates  an  impressive shadow  geometry of a black hole. Moreover,  it has been suggested that  the gravitational lensing is the  feature   distinguishing  the Newtonian and  the general-relativistic cases. This can  enhance non only  the bending of  the light   but also  it underlines  the black hole shadow.
%The shadow takes shape through a critical curve, dividing capture and scattering orbits. It encompasses a geometrically thick, optically thin region filled with emitters spiraling into or away from the black hole, associated with a distant, homogeneous, isotropic emission ring. The shadow's unique visual signature stems from its intensity dip, perfectly coinciding with the dark region, making it readily observable.
Its size  depends mainly on  the black hole  intrinsic parameters.  Its  contour, however,  remains dynamic due to the instability of  the light rays from the photon sphere. To distant observers, the shadow appears as a dark. This is a two-dimensional disk, being  beautifully set against its bright with  uniform surroundings.

In this section, we are interested in spherically free-falling accretion around  the  Dunkl black hole, originating from an infinite distance. The present  investigation  is  based on  the approach proposed in \cite{Jaroszynski:1997bw,Bambi:2012tg,Bambi:2013nla}. Inspired by such a work,  we provide a  realistic representation of the accretion disk  shadow.  %{\bf However, it is important to  note that,  in reality, the actual image of the black hole remains elusive, without visible limit in the immensity of the universe.  As a result, it is impossible to use a static accretion disk model, since the dynamic nature of the accretion disk encircling the black hole is an integral part of the process. Indeed, the accretion disk produces synchrotron emissions.}

\begin{figure}[!ht]
\centering 
\includegraphics[width=0.8\textwidth]{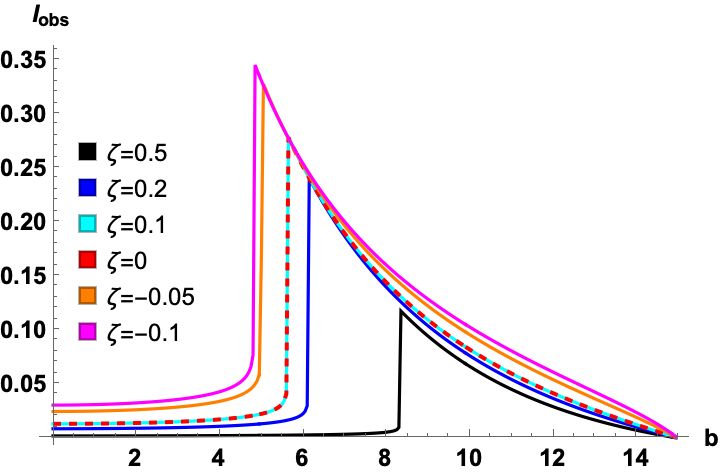}
\caption{\label{fig5}  \footnotesize\it  , Observed specific intensity at spatial infinity with $M=1$ in terms of the impact parameter, for different power parameter parameter values $\zeta$  }
\end{figure}

Roughly, we  would like to inspect  the impact of the  Dunkl parameter  $\zeta$ on the spacetime structure and  the observational characteristics.
Sending  the observer to the infinity, the  observation specific intensity  (measured usually in $\rm erg s^{-1} cm^{-2} str^{-1} Hz^{-1}$) can be determined using the techniques developed in \cite{Jaroszynski:1997bw,Bambi:2013nla}.  Indeed,  it is expressed as follows 
\begin{align}
{I}_{obs}(\nu_{obs},b)=\int _{\Gamma }g^3 j( \nu _e) d\ell_{{prop}}, \label{Eq4.1}
\end{align}
where $g$ represent the redshift factor. It denoted that  $\nu _e$  and  $\nu _{obs}$  are  the photon frequency and the   observed photon frequency, respectively,   $b$ stands for the impact parameter. Considering the emitter  in the rest frame,  $j( \nu _e)\propto  \frac{\delta \left(\nu _e-\nu _f\right)}{r^2}$ is the emissivity per unit volume,  proportional to the  $1 / r^2$ radial profile \cite{Bambi:2013nla}  via the     delta function $\delta$.  $\nu_f$ is the light radiation   frequency  being  assumed to be monochromatic.  $d\ell_{{prop}}$   denotes  the infinitesimal proper length. Taking a  black hole geometry, the redshift factor can be written as
\begin{equation}
g=\frac{\hat{\mathcal{K}}_{\rho } u _0^{\rho }}{\hat{\mathcal{K}}_{\sigma } u _e^{\sigma }}, \quad  \hat{\mathcal{K}}^{\mu }=\dot{x}_{\mu },   \label{EQ4.8}
\end{equation}
where $\hat{\mathcal{K}}^{\mu }$  indicates  the four-velocity of the photon.  $u_0^{\mu }$  is the four-velocity of the static observer being identified with $=(1,0,0,0)$.  However, the quantity $u_e^{\mu }$  indicates  the four-velocity of the  infalling accretion involving the form
\begin{equation}
u_e^t={f (r)}^{-1},\quad u_e^r=-\sqrt{{1-f (r)}}, \quad u_e^{\theta }=u_e^{\varphi }=0.   \label{EQ4.9}
\end{equation}
Taking a  null geodesic, we  could  extract the photon  four-velocity. Indeed, it is given by 
\begin{equation}
\hat{\mathcal{K}}_t=\frac{1}{b}, \quad \frac{\hat{\mathcal{K}}_r}{\hat{\mathcal{K}}_t}= \pm \frac{1}{f (r)}\sqrt{1-f(r) \frac{b^2}{r^2}}.  \label{EQ4.10}
\end{equation}
In this way,  the sign $\pm$   characterize   the situation where  the photon trajectory is either towards or away from the black hole. Thus, the redshift factor of the infalling accretion enhances
\begin{equation}
g=\left(u_e^t+\left(\frac{\hat{\mathcal{K}}_r }{\hat{\mathcal{K}}_e}\right)u_e^r \right) ^{-1}. \label{EQ3.9}
\end{equation}
On the other hand, the  proper distance should  take the form 
\begin{equation}
d\ell_{prop}=\hat{\mathcal{K}}_\mu u_e ^\mu d\lambda=\frac{\hat{\mathcal{K}}_t}{g |\hat{\mathcal{K}}_r|} d r. \label{EQ3.10}
\end{equation}
Integrating Eq.\eqref{Eq4.1} over all possible  frequencies, we could obtain  the  observation intensity for the  infalling spherical accretion. The computation leads to 
\begin{equation}
I_{{obs}}\propto \int _{\Gamma }\frac{\mathcal{G}_i^3  \hat{\mathcal{K}}_t dr}{r^2 |\hat{\mathcal{K}}_r|}.\label{EQ3.11}
\end{equation}
Using  the above equations, we explore the shadow and brightness of the power Dunkel black hole,  surrounded by the infalling spherical accretion.
In Fig.\ref{fig5}, we  depict  the luminosity distribution by considering   different values of the   parameter $\zeta$.

From  this  figure, we can observe that the specific intensity $I_{obs}$   augments  stylishly with    the impact parameter $b$ and  then it reaches  a  peak value in the critical case $b\sim b_{ph}$. This  universal behavior has appeared   for  different values of the Dunkl parameter $\zeta$.  For  $b>b_c$, the specific intensity $I_{obs}$ involves  a trending down.  In the limit where   $b$ goes  to infinity, the observation intensity vanishes,  namely $I_{obs}\sim 0$.

\begin{figure}[!ht]
\centering
			\begin{tabbing}
			\centering
			\hspace{10.3cm}\=\kill
\includegraphics[scale=.3]{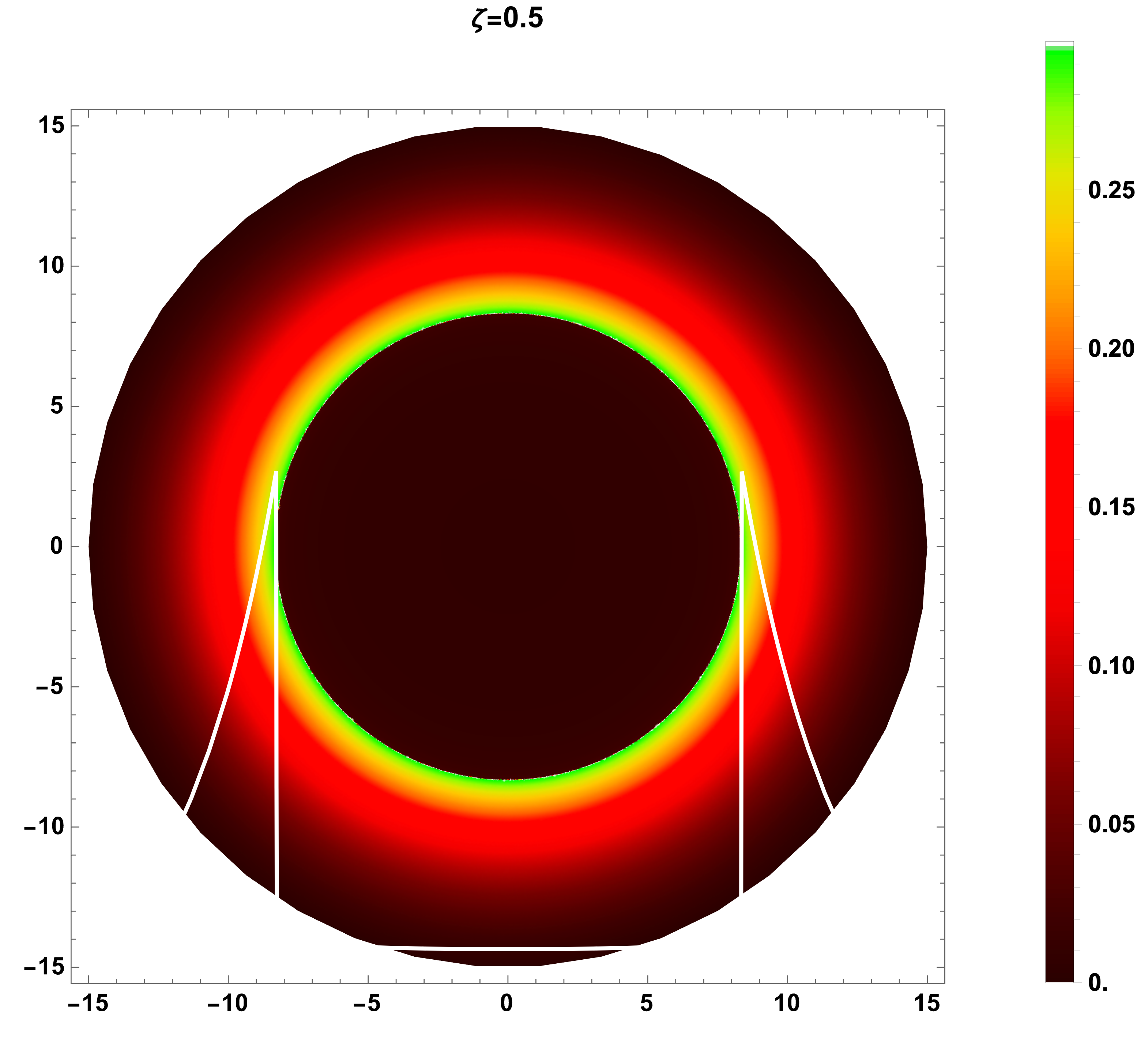}
\includegraphics[scale=.3]{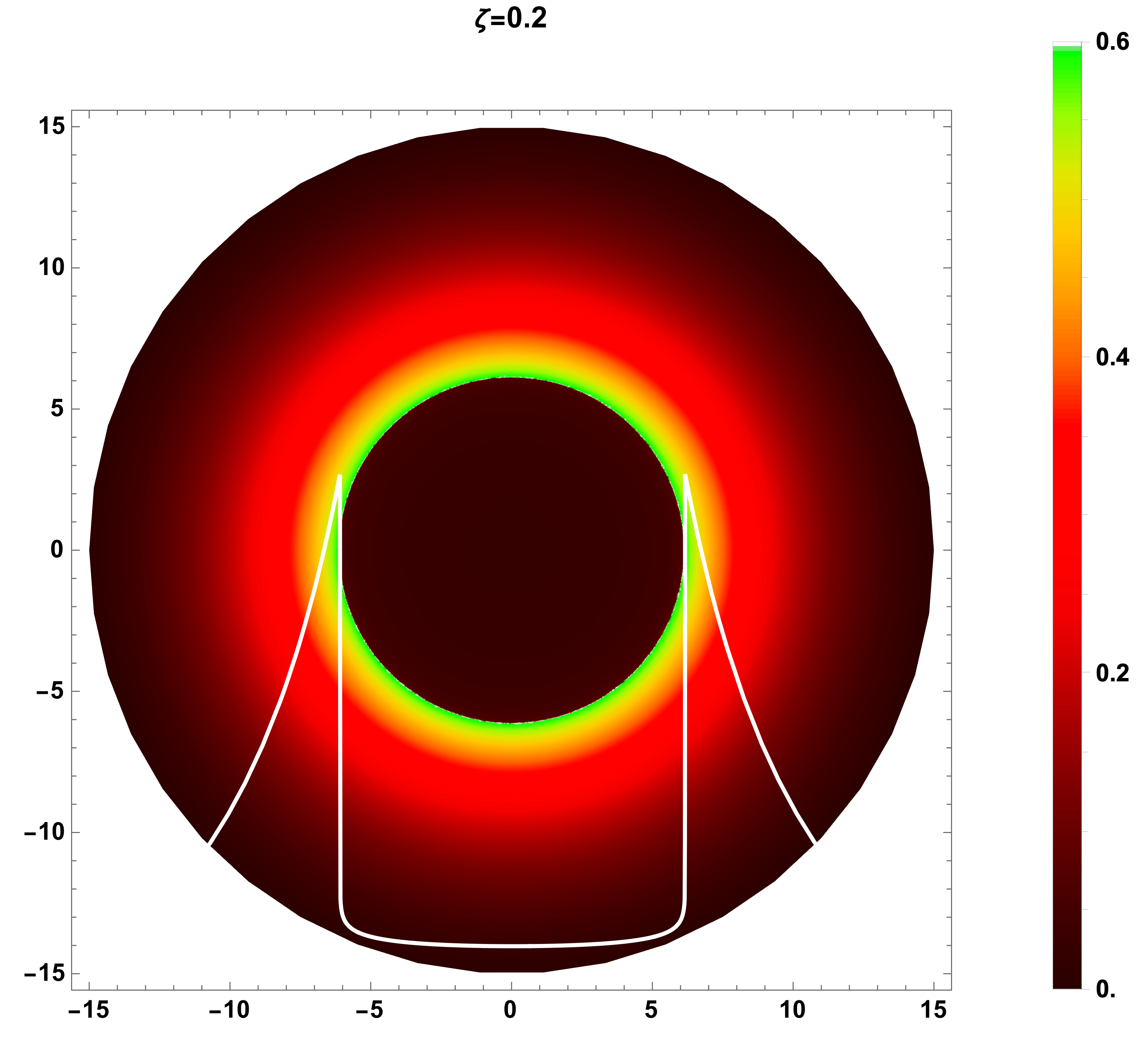}
\includegraphics[scale=.3]{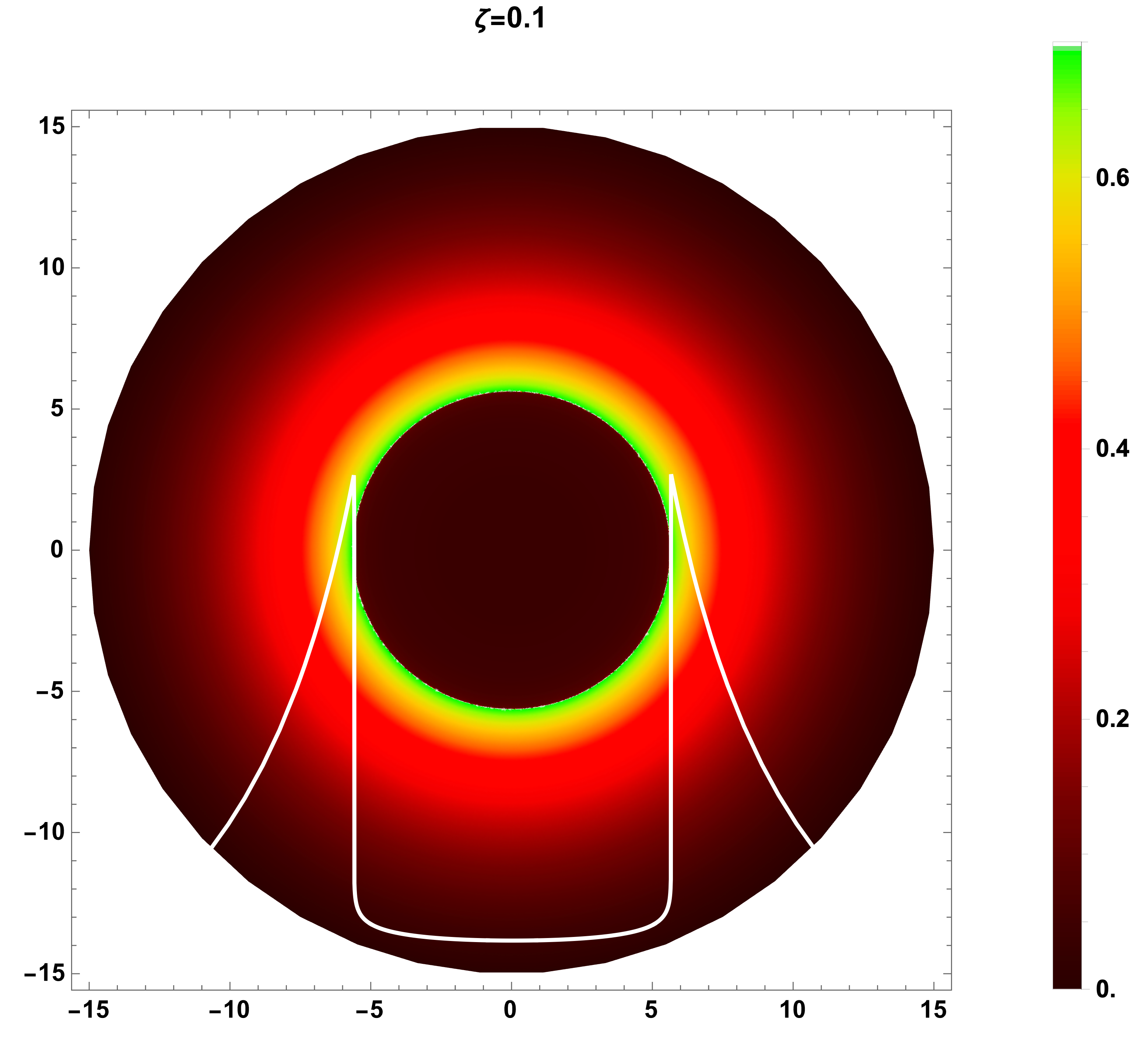}\\
\includegraphics[scale=.3]{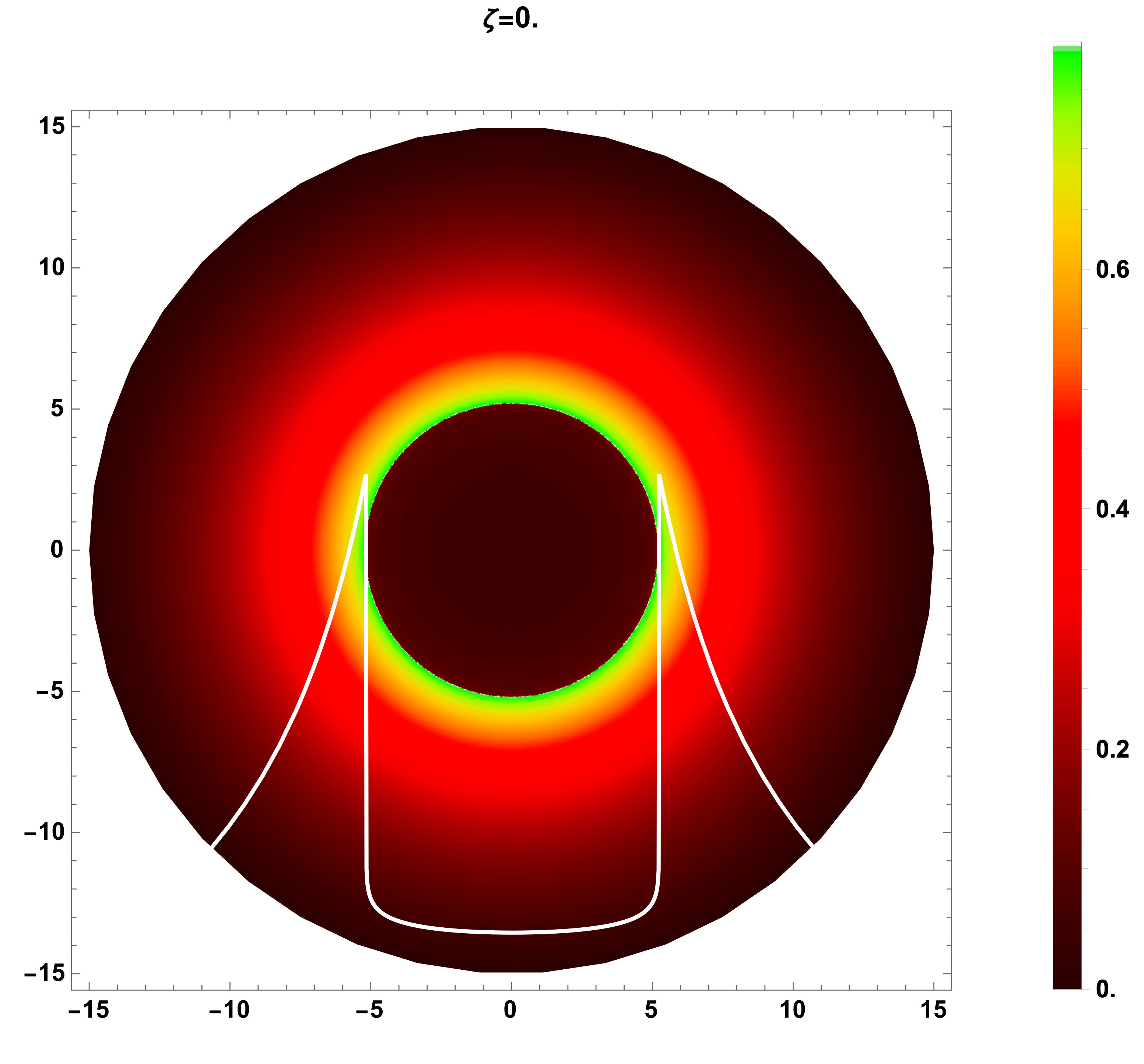}
\includegraphics[scale=.3]{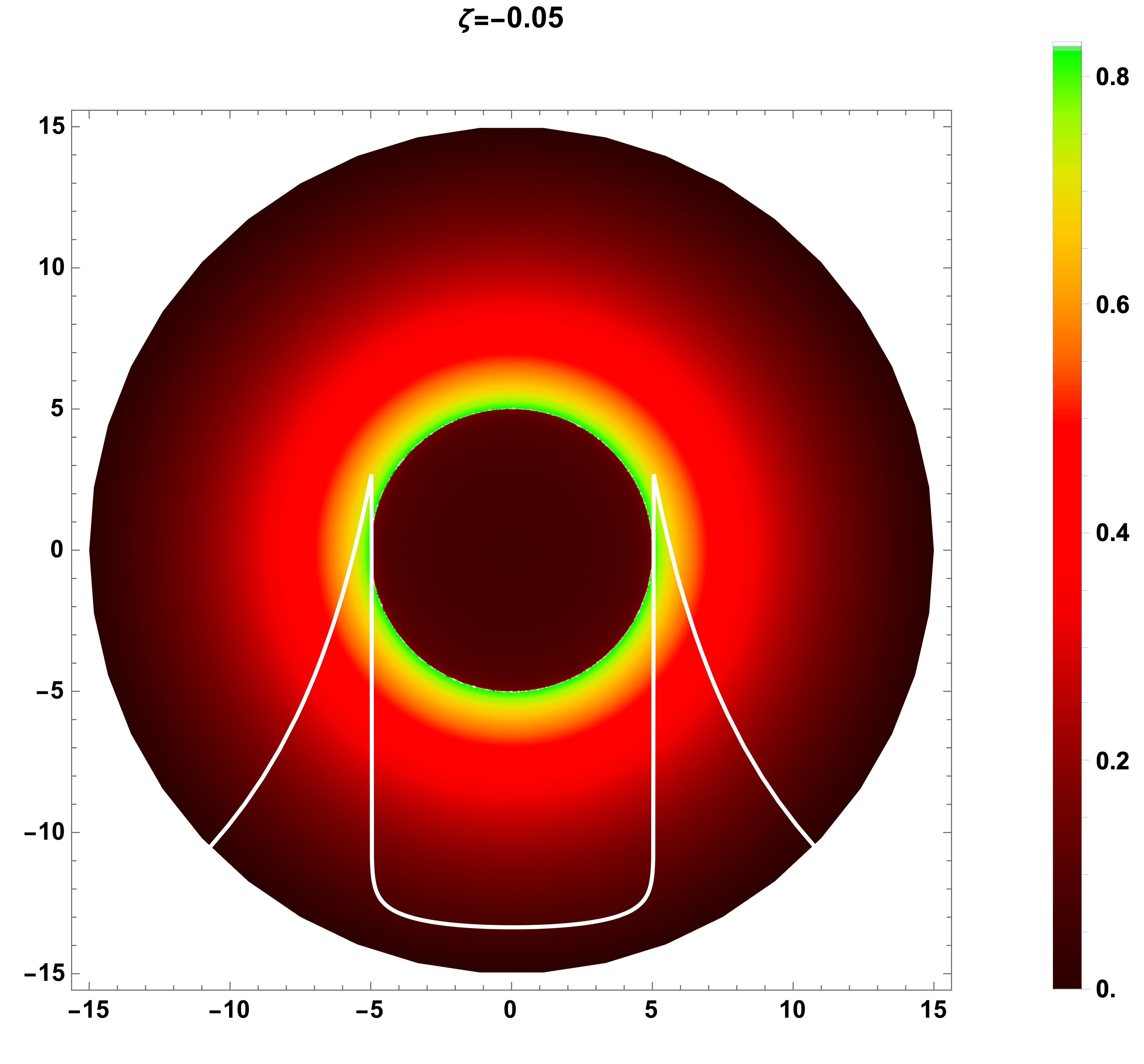}
\includegraphics[scale=.3]{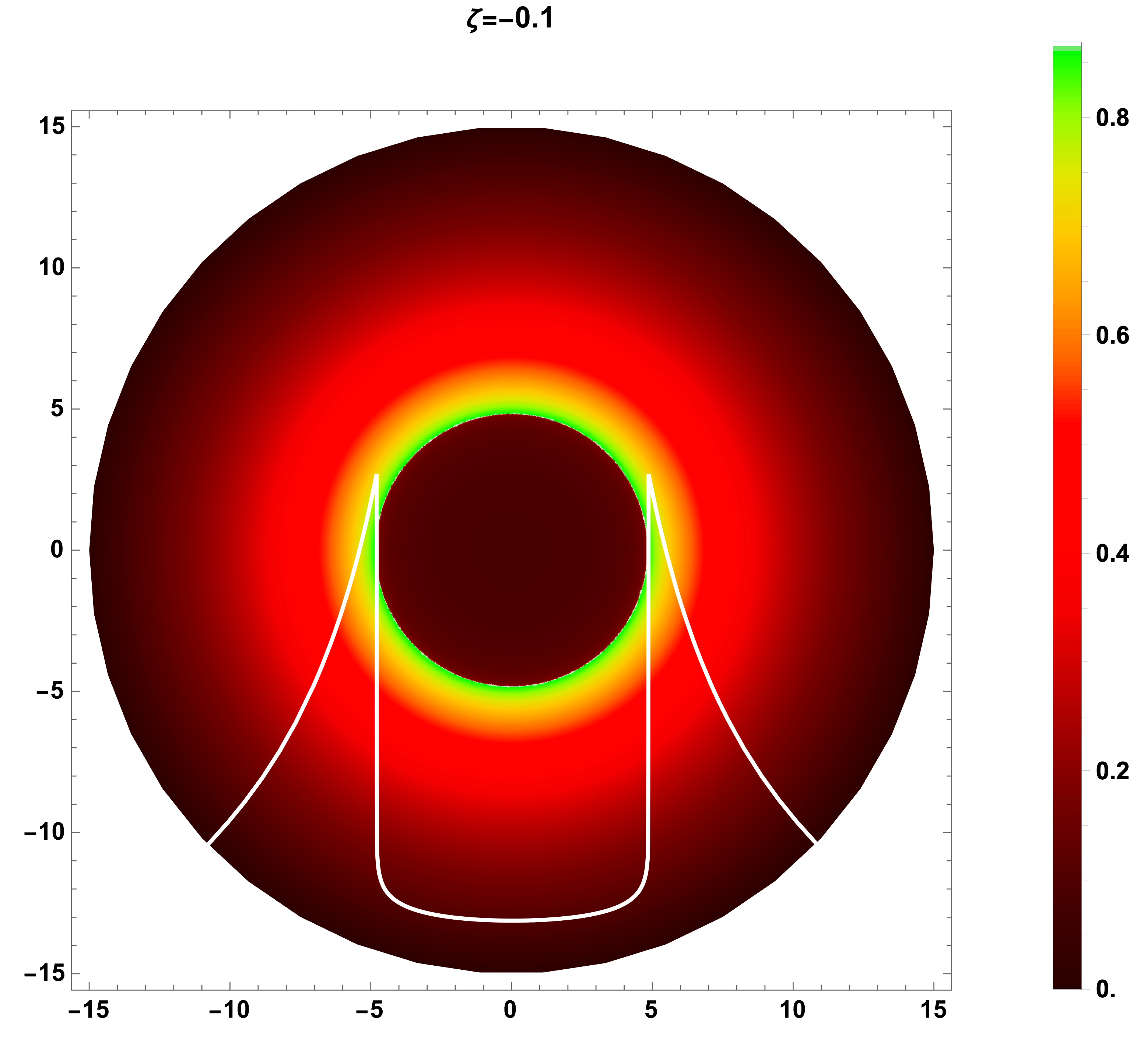}\\
		   \end{tabbing}

        \caption{\footnotesize{\it Black hole shadows with spherical accretion  by varying  Dunkel parameter $\zeta$.}}
        \label{figure6}
\end{figure}

 Fig.\ref{figure6}  presents  a two-dimensional picture  of the observed shadows of the Dunkl black hole  in terms of  $\xi$.
%\begin{figure}[!ht]
%     \begin{subfigure}[b]{0.32\textwidth}
%         \centering
%         \includegraphics[scale=.41]{SphericalBHchargefov15gamma1-2}
%         \caption{${\bm \gamma=\frac{1}{2}}$}
%         \label{fig1-1}
%     \end{subfigure}
%     \hfill
%     \begin{subfigure}[b]{0.32\textwidth}
%         \centering
%         \includegraphics[scale=.41]{SphericalBHchargefov15gamma3-4}
%         \caption{${\bm \gamma=\frac{3}{4}}$}
%         \label{fig1-2}
%     \end{subfigure}
%     \hfill
%     \begin{subfigure}[b]{0.32\textwidth}
%         \centering
%         \includegraphics[scale=.41]{SphericalBHchargefov15gamma2}
%         \caption{${\bm \gamma=2}$}
%         \label{fig1-3}
%     \end{subfigure}
%        \caption{\footnotesize{\it Images of the black hole shadow with spherical accretion  for the power law parameters $\gamma$ .}}
%        \label{figure8}
%\end{figure}

It has been observed that a bright ring with a higher luminosity is situated in close proximity to the photon sphere. Moreover, the inner region of the photon sphere is not entirely devoid of light.    It has been remarked a small luminosity near to  the photon sphere. This has been explained by   a  radiation  tiny fraction escaped   from the black hole. As anticipated, increasing the parameter $\zeta$ leads to an expansion in the black hole shadow size, while  it  simultaneously reduces  the maximum luminosity of the shadow image. This behavior is consistent with the trends observed in Fig.\ref{fig5}.

%\newpage

\subsection{Light deflection angle in vacuum  and medium backgrounds}

Now, we move to  discuss the behaviors of  the light  deflection angle  by the  Dunkl black hole by focusing on  the effect of the involved parameter. There are several methods available to compute this optical quantity. Notably, Gibbons and Werner have utilized the Gauss-Bonnet theorem \cite{Gibbons:2008rj,Arakida:2017hrm}, which connects the differential geometry of a surface to its  topology. By applying  the optical geometry, one can compute  the deflection angle of light rays passing near various spherically symmetric black hole geometries. Here, we adopt these techniques to derive the expression for the deflection angle in the weak-field limit induced by the Dunkl black hole. %Using appropriate assumptions, the Gauss-Bonnet theorem simplifies to provide a reduced integral expression for the deflection angle, given by
We begin by calculating the optical metric on the equatorial plane, being given by  $\theta=\pi/2$, using the line element \eqref{22d}. Precisely,  one finds
%Then, for the considered null geodesics for which $ds^{2}=0$, the optical metric reads as follows
\begin{equation}\label{optmet2}
dt^{2}=\frac{dr^{2}}{f(r)^2}+\frac{r^{2}}{f(r)}d\phi^{2}\,.
\end{equation}
The Gaussian optical curvature $\bar{K}=\frac{\bar{R}}{2}$,  where  $\bar{R}$ is the Ricci scalar of the optical metric \eqref{optmet2},   takes the following form 
\begin{eqnarray}\label{gauoptcur1}\nonumber
\bar{K}&=&M \left(-\frac{2}{r^3}+ \frac{4 \zeta  \log (r)}{3 r^3}-\frac{4 \zeta ^2 \log (r) (3 \log (r)+2)}{27    r^3}\right)\\
&+& M^2 \left(\frac{3}{r^4}+\frac{2 \zeta  (4-6 \log (r))}{3 r^4}+\frac{\zeta ^2 (72 (\log (r)-1) \log (r)-4)}{27 r^4}\right)+\mathcal{O}(M^3,\zeta^3).
\end{eqnarray}
To calculate the deflection angle, it is essential to consider a non-singular manifold $\mathcal{D}_{\tilde{R}}$ with a geometric scale $\tilde{R}$ to apply the Gauss-Bonnet theorem. According to \cite{Gibbons:2008rj,Arakida:2017hrm}, one has 
\begin{equation}\label{gaubon1}
\int\int_{\mathcal{D}_{\tilde{R}}}\bar{K}dS+\oint_{\partial\mathcal{D}_{\tilde{R}}}kdt+\sum_{i}\varphi_{i}=2\pi\varrho(\mathcal{D}_{\tilde{R}})\,,
\end{equation}
%where $dS=\sqrt{\bar{g}}drd\phi$ and $dt$ are the surface and line element of the optical metric Eq.\eqref{optmet2}, respectively, $\bar{g}$ is the determinant of the optical metric, $k$ denotes the geodesic curvature of $\mathcal{D}_{\tilde{R}}$, and $\varphi_{i}$ is the jump (exterior) angle at the $i$-th vertex, and also, $\zeta(\mathcal{D}_{\tilde{R}})$ is the Euler characteristic number of $\mathcal{D}_{\tilde{R}}$. One can set $\zeta(\mathcal{D}_{\tilde{R}})=1$. Then, considering a smooth curve $y$, which has the tangent vector $\dot{y}$ and acceleration vector $\ddot{y}$, the geodesic curvature $k$ of $y$ can be defined as follows where the unit speed condition $\tilde{g}\left(\dot{y},\dot{y}\right)=1$ is employed
%------------------------------
where  one used $dS=\sqrt{\bar{g}}drd\phi$ and $dt$ represent the surface and line elements of the optical metric in Eq.\eqref{optmet2}, respectively. Here, $\bar{g}$ is the determinant of the optical metric.  $k$  is the geodesic curvature of $\mathcal{D}_{\tilde{R}}$.  $\varphi_{i}$ indicates the exterior angle at the $i$-th vertex. Additionally, $\varrho(\mathcal{D}_{\tilde{R}})$ represents the Euler characteristic of $\mathcal{D}_{\tilde{R}}$, which can be set to $\varrho(\mathcal{D}_{\tilde{R}}) = 1.$  For a smooth curve $y$ with a tangent vector $\dot{y}$ and an acceleration vector $\ddot{y}$,  we can define the geodesic curvature $k$ of $y$  via the  speed constraint $\tilde{g}\left(\dot{y}, \dot{y}\right) = 1$. Indeed, it is given by 
\begin{equation}\label{geocur}
k=\tilde{g}\left(\nabla_{\dot{y}\dot{y},\ddot{y}}\right),
\end{equation}
which measures  the deviations of $y$ from  a geodesic. In the limit where  $\tilde{R} \rightarrow \infty$, the two jump angles $\varphi_{s}$ (at the source) and $\varphi_{o}$ (at the observer) approach $\pi/2$, leading to $\varphi_{s} + \varphi_{o} \rightarrow \pi$. For the curve $C_{\tilde{R}} := r(\phi)$, the geodesic curvature is $k(C_{\tilde{R}}) = |\nabla_{\dot{C}{\tilde{R}}}\dot{C}{\tilde{R}}| ,,, \myeqq ,,, 1/\tilde{R}$, and we obtain $\lim_{\tilde{R} \rightarrow \infty} dt = \tilde{R} d\phi$. Thus,  one has $k(C_{\tilde{R}})dt = d\phi$. In fact, the Gauss-Bonnet theorem simplifies to the  form
\begin{equation}\label{gaubon2}
\int\int_{\mathcal{D}_{\tilde{R}}}\bar{K}dS+\oint_{C_{\tilde{R}}}kdt\,\,\,\myeq\,\,\,\int\int_{\mathcal{D}_{\infty}}\bar{K}dS+\int_{0}^{\pi+\Theta}d\phi=\pi\,.
\end{equation}
Exploiting techniques for calculating the deflection angle (see Refs \cite{Gibbons:2008rj,Arakida:2017hrm} and references therein), one gets 
\begin{equation}\label{defang}
\Theta=\pi-\int_{0}^{\pi+\Theta}d\phi=-\int_{0}^{\pi}\int_{\frac{\xi}{\sin\phi}}^{\infty}\bar{K}dS\,.
\end{equation}
Combining the above equations, the Gaussian optical curvature for the Dunkel dark compact object is found to be 
\begin{eqnarray}\label{kstvg}
\bar{K}&\approx&  M \left(-\frac{2}{r^2}
   +\zeta  \left(\frac{4 \log (r)}{3
   r^2}-\frac{3}{r^2}\right)+\zeta ^2 \left(-\frac{3}{4 r^2}-\frac{4 \log ^2(r)}{9 r^2}+\frac{46 \log (r)}{27
   r^2}\right)
   \right)\\ \nonumber
   &+& M^2 \left(-\frac{3}{r^3}+\zeta  \left(\frac{4 \log (r)}{r^3}-\frac{47}{6
   r^3}\right)+\zeta ^2 \left(-\frac{1355}{216 r^3}-\frac{8 \log ^2(r)}{3 r^3}+\frac{86 \log
   (r)}{9 r^3}\right)\right)+\mathcal{O}(M^3,\zeta^3).
%\Big]\bigg\}
\end{eqnarray}
Moreover, the  optical metric \eqref{optmet2} for the Dunkl dark compact object, in accordance with the metric coefficients \eqref{22d}, can be approximated as
\begin{equation}\label{surele1}
dS=\sqrt{\bar{g}}\,drd\phi=\frac{r}{h(r)\sqrt{f(r)}}dtd\phi\approx rdrd\phi\,.
\end{equation}
In this way, the deflection angle for the Dunkl  black hole can be  expressed  as follows
\begin{equation}\label{defangstvg1}
\begin{split}
\Theta & =-\int_{0}^{\pi}\int_{\frac{\xi}{\sin\phi}}^{\infty}\bar{K}dS.
\end{split}
\end{equation}
After computations, this quantity is found to be 
\begin{equation}\label{defangstvg}
\begin{split}
\Theta & 
\approx%M\Big(\frac{4}{b}+\zeta\Big(\frac{2}{3b}+\frac{2\log(16)-8\log(b)}{3b}\Big)\Big)\\&+\frac{1}{54 b}-\frac{2 \pi ^2}{27 b}+\frac{8 \log ^2(b)}{9 b}+\frac{4 \log ^2(2)}{9   b}+\frac{4 \log (b)}{27 b}-\frac{16 \log (2) \log (b)}{9 b}-\frac{\log (256)}{54 b}
M\Big\{\frac{4}{b}+\frac{\zeta}{3b}\Big[2+\log (256)-8 \log (b))\Big]+\frac{\zeta^2}{54b}\Big[48 \log ^2(b)+(8-96 \log (2)) \log (b)\\
&-4 \pi ^2+1+8 \log (2) (\log (8)-1)\Big]\Big\}+M^2\Big\{\frac{3\pi}{4b^2}+ \frac{\zeta\pi}{24b^2}\Big[47-12\log(4)-24\log(b) \Big]\\
&+\frac{\zeta^2\pi}{864 b^2}\Big[48 \log (b) (12 \log (4 b)-43)+48 \pi ^2+1067+24 \log (4) (6 \log (4)-43)\Big]\Big\}+\mathcal{O}(M^3,\zeta^3)
\end{split}
\end{equation}

\begin{figure}[!ht]
\centering 
\includegraphics[width=0.6\textwidth]{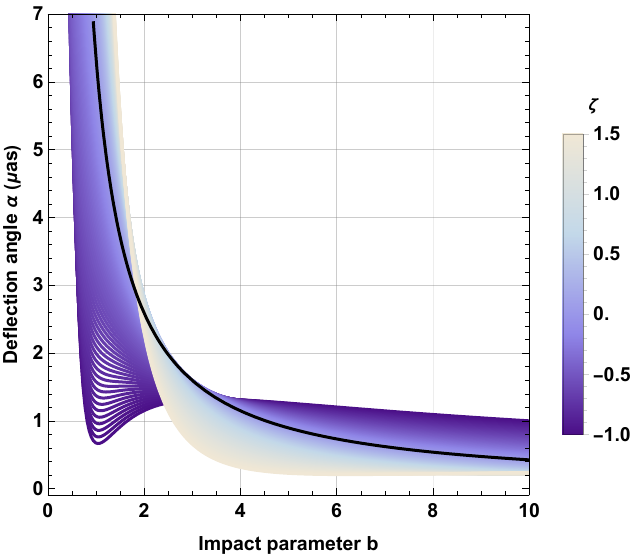}
\caption{\label{figdef}  \footnotesize\it   Observed specific intensity at spatial infinity with $M=1$ in terms of the impact parameter by taking  certain  parameter values $\zeta$.  }
\end{figure}

Having expressed  the deflection angle for such a black hole  in the vacuum medium, we move to illustrate graphically the effect of  the Dunkl parameter on this quantity. Indeed,  we depict in Fig.\ref{figdef} the light  deflection
angle $\Theta$  in terms of  the impact parameter $b$.

From Fig.\ref{figdef},  we  can  observe  that the deflection angle behavior shows a minimum for $\zeta\leq-0.8$. Then, it  increases and after that  it  decreases gradually and goes to positive infinity. As  $\zeta$ grows  the minimum point,  it  disappears and  only  the decreasing behaviors persists.
%\textcolor{red}{ %Our current analysis focuses on homogeneous and non-homogeneous plasma profiles and observing their effect on the shadows. In this section, we will briefly discuss the impact of the plasma medium on the shadow radius following [60]
%--- V. Perlick, O. Y. Tsupko, and G. S. Bisnovatyi-Kogan, Phys. Rev. D 92, 104031 (2015),
%arXiv:1507.04217 [gr-qc].
%----. These findings will be used to analyze the effects of the above mentioned plasma profiles on generalized Hayward spacetimes.
%We would restrict our analysis to the non-magnetized cold plasma with the plasma frequency of the form,}
In reality, the  astrophysical objects are typically enveloped by an interstellar medium, often modeled as plasma in the  event horizon vicinity. As photons traverse this plasma, their interactions with the medium modify their trajectories. This interaction, in turn, influences the apparent size of the central   shadow as seen by a distant observer.

To incorporate plasma effects, we assume that the photon travels from a vacuum into a hot, ionized gas medium, where $v$ is the  light speed  in the plasma. Indeed,  the refractive index, $n(r)$, can then be expressed as follows 
\begin{equation}
n_\text{Plasma}(r)\equiv\frac{v}{c}=\frac{1}{dr/dt},\quad\quad \{\because c=1\}.
\end{equation}
 Following \cite{Perlick:2015vta,Crisnejo:2019ril},  the refractive index $n(r)$ for the associated  black hole  reads as 
\begin{equation}
n_\text{Plasma}(r)=\sqrt{1-\frac{\omega_e^2}{\omega_\infty^2}f(r)},
\end{equation}
where $\omega_e$ denotes the electron plasma frequency. $\omega_\infty$ represents  the photon frequency as observed at infinity. Consequently, the line element in Eq.\eqref{22d} transforms into the following form
\begin{equation}\label{nr}
dt^2=g_{ij}^{opt}dx^idx^j=n_\text{Plasma}^2(r)\left[\frac{dr^2}{f(r)^2}+\frac{r^2d\phi}{f(r)}\right].
\end{equation}
The deflection angle, originally derived under vacuum conditions,  could  take  a new form influenced by the plasma properties. Indeed,  it can be expressed as follows
\begin{eqnarray}\nonumber
\Theta_{\text{Plasma}}&\approx&\left\{\left(\frac{4}{b}+\frac{2\omega_e}{b \omega_\infty}+\frac{2\omega_e^2}{b \omega_\infty^2} \right)+\left(\frac{2 (-4 \log (b)+1+\log (16))}{3 b}\right.\right.\\ \nonumber
&+& \left.\left. \frac{2 (-4 \log (b)+1+\log (16))}{3 b}+
\frac{\omega_e^2 (-4 \log (b)-9+\log (16))}{3 b \omega_\infty^2}
\right)\zeta
\right\}M\\ \nonumber
&+& 
\left\{\left(\frac{3 \pi }{4 b^2}-\frac{\pi \omega_e}{2 \left(b^2 \omega_\infty\right)}-\frac{3 \pi \omega_e^2}{2 b^2 \omega_\infty^2}
   \right)\right.+\left(
 \frac{-24 \pi  \log (b)+47 \pi -12 \pi  \log (4)}{24 b^2}\right. \\ \nonumber
&+&
\left. \left. \frac{\omega_e (8 \pi  \log (b)-11 \pi +\pi  \log (256))}{12 b^2 \omega_\infty}+\frac{\omega_e^2 (8 \pi  \log (b)-5 \pi +\pi  \log (256))}{4 b^2 \omega_\infty^2}\right.\right.\\
&+&\left.\left.
\frac{\omega_e^2 (8 \pi  \log (b)-5 \pi +\pi  \log (256))}{4 b^2 \omega_\infty^2} \right) \zeta
   \right\}M^2+\mathcal{O}(M^3,\zeta^2,\omega_e^3).
\end{eqnarray}
The presence of a medium introduces significant dispersion effects on  the light rays instead of propagation in a vacuum. { One  possibility of such a  medium, consistently found in astrophysical contexts, is dark matter, which often forms vast halos around black holes, as observed by  EHT \cite{Xu:2018wow,Chen:2024mlr,Wu:2024hxr}.  It has been remarked that  the Dark matter  has the  ability to enshroud entire galaxies and permeate both interstellar and intergalactic spaces being a    central  point of certain  scientific investigations\cite{Rahman:2023sof,Pantig:2024rmr,Kazempour:2024lcx}}.  Combining  the dark matter  with black holes  provides  essential insights into gravitational lensing phenomena. To assess the impact of dark matter on the deflection angle of light, we derive the refractive index based on scatterers in the medium, as detailed in references\cite{Ovgun:2018oxk,Latimer:2013rja,Ovgun:2020yuv}. Indeed, we have 
\begin{equation}\label{nnr}
n_\text{DM}=1+\mathcal{B}u+v w^2,
\end{equation}
where  $w$ denotes the light  frequency. The quantity  $\mathcal{B}$ is identified with  $=\rho_0/4mw^2$  where $\rho_0$ is the mass density of the  dark matter particles, and $m$ is their mass. The parameter $u=-2\epsilon^2e^2$ is defined. It  is denoted that  $\epsilon$ is  the charge of the scattered matter  in units of $e$.  $v$ represents  a non-negative value. Higher  order $\mathcal{O}(w^2)$ and  account for the polarizability of the dark matter particle could  define  the refractive index for an optically inactive medium. Specifically, the $w^{-2}$ term is associated with  a charged dark matter candidate, while the $w^{2}$ term corresponds to a neutral candidate. It is worth noting that, a linear term in $w$ may appear once one has   the parity and the charge parity asymmetries.

Recalling  the optical metric Eq.\eqref{nr} with the optical index Eq.\eqref{nnr}, the deflection angle associated with the Dunkel black hole 
enveloped by dark matter can be given by 
\begin{equation}
\Theta_{\text{DM}}=\frac{\Theta}{\left(1+\mathcal{B}u+v w^2\right)^2}.
\end{equation}

The analysis reveals that the deflection angle rises by  increasing values of the parameter $w$ within the weak field regime. This trend suggests that heightened dark matter activity intensifies lensing profile distortions. While the dynamics of dark matter near black holes are not as well understood as its distribution in outer-galactic regions, its essential role in enhancing both the distortions and the resulting deflection angle remains a consistent and significant factor.

 %\textcolor{red}{ experimental constraints on deflection angle}
\section{Constraints from EHT observations of $\textrm{M87}^\star$ and $\textrm{Sgr~A}^\star$ black holes}
In this section, we aim to compare the shadow results obtained from our analysis with the observations of  the supermassive black holes captured by EHT. This comparison   constrains the Dunkl parameter $\zeta$ by utilizing the observational  data of $\textrm{M87}^\star$ and $\textrm{Sgr~A}^\star$ provided by the EHT international  collaboration.

Specifically, we focus on  the key observables related to the  black hole shadows: the angular diameter $\mathcal{D}$ and the deviation from the Schwarzschild geometry ${\bm \delta}$. By matching our theoretical predictions with the observed data for $\textrm{M87}^\star$ and $\textrm{Sgr~A}^\star$, we  wouyld like to  establish tighter constraints on the Dunkl parameter.  According to  \cite{Banerjee:2019nnj}, the angular diameter $\mathcal{D}$ is a crucial observable defined as
 \begin{align}
\mathcal{D}=\frac{G m}{c^2}\left(\frac{\Delta Y }{d}\right),
\label{Eq1}
\end{align}
where $d$ denotes  the distance from Earth to $\textrm{M87}^\star$ or $\textrm{Sgr~A}^\star$.  Moreover,  $\Delta Y$ represents the maximum length of the shadow along the direction $Y$   in the $(X,Y)$ plane,  being orthogonal to  the  $X$ axis. The EHT collaboration provides  the deviation parameter ${\bm \delta}$ to quantify the difference between the infrared shadow radius. For the Schwarzschild black hole \cite{EventHorizonTelescope:2022apq}, it takes the form 
 \begin{equation}\label{ddelta}
{\bm \delta}=\frac{r_\text{s}}{r_\text{Sch}}-1.
\end{equation}

According to EHT collaboration \cite{EventHorizonTelescope:2019dse,EventHorizonTelescope:2019pgp,EventHorizonTelescope:2019ggy}, the observed parameters of the supermassive black hole $\textrm{M87}^\star$ at the center of the $\textrm{M}87$ galaxy include an angular diameter of $\mathcal{D} = 42 \pm 3 \mu$as and a fractional deviation ${\bm \delta}$ from the predicted Schwarzschild black hole shadow diameter reported by the EHT \cite{EventHorizonTelescope:2022xqj}. The latter  remains consistent regardless of the telescope, image processing techniques, or  used simulations.  Based on the black hole image of $\textrm{M87}^\star$ \cite{EventHorizonTelescope:2021dqv}, detailed values for ${\bm \delta}$ are provided in Tab.\ref{m87_bounds} 
\begin{table}[!ht]
	\centering
	\begin{tabular}{lccr} % four columns, alignment for each
	& \multicolumn{3}{c}{$\textrm{M}87^\star$ estimates}\\[1mm]\\[-3.5mm]
		\hline\hline
		 & Deviation ${\bm \delta}$ & $\text{1-}\sigma$ bounds & $\text{2-}\sigma$ bounds\\
		\hline
		EHT & $-0.01^{+0.17}_{-0.17}$ & $4.26\le \frac{r_{\text{sh}}}{M}\le 6.03$ &  $3.38\le \frac{r_{\text{sh}}}{M}\le 6.91$\\
		\hline
	\end{tabular}\vspace{.5cm}
	\begin{tabular}{ c c | c c c }
    
        & \multicolumn{4}{c}{Sgr $A^*$ estimates}\\[1mm]\hline\hline\\[-3.5mm]
        
        & &\hspace{\tabley} Deviation $\delta$ &\hspace{\tablex} 1-$\sigma$ bounds &\hspace{\tablex} 2-$\sigma$ bounds\\[1mm]\hline
        
        \parbox[ht]{3mm}{\multirow{2}{*}{\rotatebox[origin=c]{90}{\textit{\scriptsize eht-img}\hspace{1mm}}}} & {VLTI}\hspace{\tabley} &\hspace{\tabley} $-0.08^{+0.09}_{-0.09}$ &\hspace{\tablex} {$4.31\le \frac{r_{\text{sh}}}{M}\le 5.25$} &\hspace{\tablex} {$3.85\le \frac{r_{\text{sh}}}{M}\le 5.72$}\\[1mm]
        
        & Keck\hspace{\tabley} &\hspace{\tabley} $-0.04^{+0.09}_{-0.10}$ &\hspace{\tablex} {$4.47\le \frac{r_{\text{sh}}}{M}\le 5.46$} &\hspace{\tablex} {$3.95\le \frac{r_{\text{sh}}}{M}\le 5.92$}\\[1mm]

        \hline

        \parbox[ht]{3mm}{\multirow{2}{*}{\rotatebox[origin=c]{90}{\textit{\scriptsize mG-ring}\hspace{1mm}}}} & VLTI\hspace{\tabley} &\hspace{\tabley} $-0.17^{+0.11}_{-0.10}$ &\hspace{\tablex} $3.79\le \frac{r_{\text{sh}}}{M}\le 4.88$ &\hspace{\tablex} $3.27\le \frac{r_{\text{sh}}}{M}\le 5.46$\\[1mm]
        
        & Keck\hspace{\tabley} &\hspace{\tabley} $-0.13^{+0.11}_{-0.11}$ &\hspace{\tablex} $3.95\le \frac{r_{\text{sh}}}{M}\le 5.09$ &\hspace{\tablex} $3.38\le \frac{r_{\text{sh}}}{M}\le 5.66$\\[1mm]
        
        \hline
        
    \end{tabular}

	\caption{\footnotesize\it {\bf Top: } $\textrm{M}87^\star$  bounds on the deviation parameter $\bm\delta$ and dimensionless shadow radius $r_\text{sh}/M$. {\bf Bottom:} Sagittarius A* bounds on the deviation parameter $\delta$ and shadow radius\cite{Antoniou:2022dre}.}
	\label{m87_bounds}
\end{table}

For $\textrm{M87}^\star$, the central value of ${\bm \delta}$ tends toward zero, but uncertainties in mass and distance measurements contribute to larger error margins. Utilizing the value of ${\bm \delta}$ from Eq. \eqref{ddelta} and Tab.\ref{m87_bounds}, one can impose constraints on the  black hole parameters by considering the dimensionless quantity $r_\text{s}/M$. Additionally, an inclination angle of $17^\circ$ is predicted, corresponding to the angle formed between the jet axis and the observer line of sight, assuming the rotation axis aligns with the jet. The EHT collaboration estimates the distance to $\textrm{M87}^\star$ to be $d = 16.8 \pm 0.8$ Mpc, with the mass calculated as $m = (6.5 \pm 0.7) \times 10^{9} M_{\odot}$. The angular diameter of the black hole  shadow is estimated at $42 \pm 3 \mu$as, with a $10\%$ correction applied to account for the offset between the image and the shadow, yielding a minimum shadow diameter of $37.8 \pm 2.7 \mu$as.

For $\textrm{Sgr~A}^\star$, the observed emission ring diameter is reported as $51.8 \pm 2.3 \mu$as by the EHT collaboration \cite{EventHorizonTelescope:2022wkp}, while the estimated shadow diameter is $48.7 \pm 7 \mu$as. Multiple teams have measured the mass and distance to $\textrm{Sgr~A}^\star$. Data from the Keck Observatory show a distance of $d = (7959 \pm 59 \pm 32)$ pc and a mass of $m = (3.975 \pm 0.058 \pm 0.026) \times 10^{6} M_{\odot}$, where the redshift parameter is treated as free. Assuming a redshift of unity gives a distance of $d = (7935 \pm 50)$ pc and a mass of $m = (3.951 \pm 0.047) \times 10^{6} M_{\odot}$ \cite{Do:2019txf}. Similarly, the GRAVITY collaboration from the Very Large Telescope Interferometer (VLTI) reports a mass of $m = (4.261 \pm 0.012) \times 10^{6} M_{\odot}$ and a distance of $d = (8246.7 \pm 9.3)$ pc \cite{GRAVITY:2021xju,GRAVITY:2020gka}. When accounting for optical aberrations, these measurements suggest a mass of $m = (4.297 \pm 0.012 \pm 0.040) \times 10^{6} M_{\odot}$ and a distance of $d = 8277 \pm 9 \pm 33$ pc. Furthermore, numerical models suggest that the inclination angle $i$ of $\textrm{Sgr~A}^\star$ is greater than $50^\circ$. Our analysis uses an inclination value of $i \simeq 134^\circ$ (or equivalently $46^\circ$) \cite{refId0}. Using the \textit{eht-img} algorithm \cite{EventHorizonTelescope:2022xqj},  the fractional deviation $\bm{\delta}$ and the dimensionless shadow radius $r_\text{s}/M$ have been computed. They are presented at the bottom of Tab.\ref{m87_bounds}.

\begin{figure}[!ht]
\centering
\begin{tabbing}
			\centering
			\hspace{9cm}\=\kill

\centering \includegraphics[width=.53\textwidth]{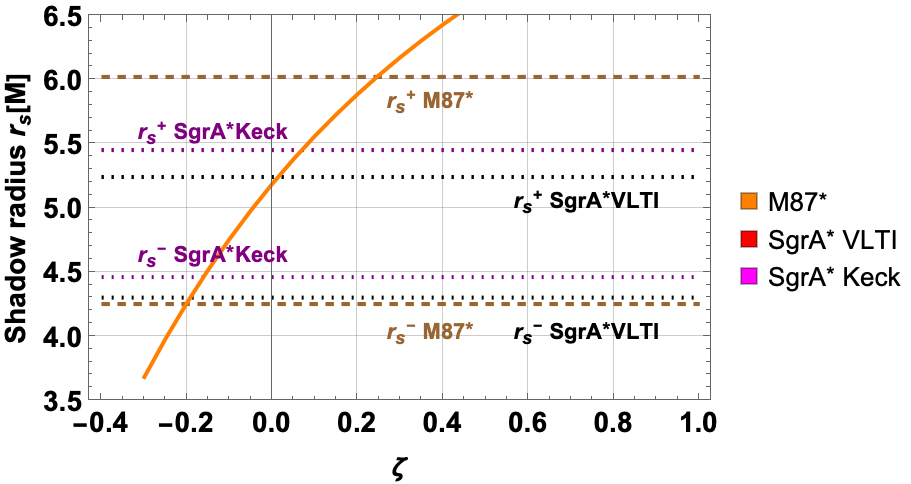}\>
\centering \includegraphics[width=.5\textwidth]{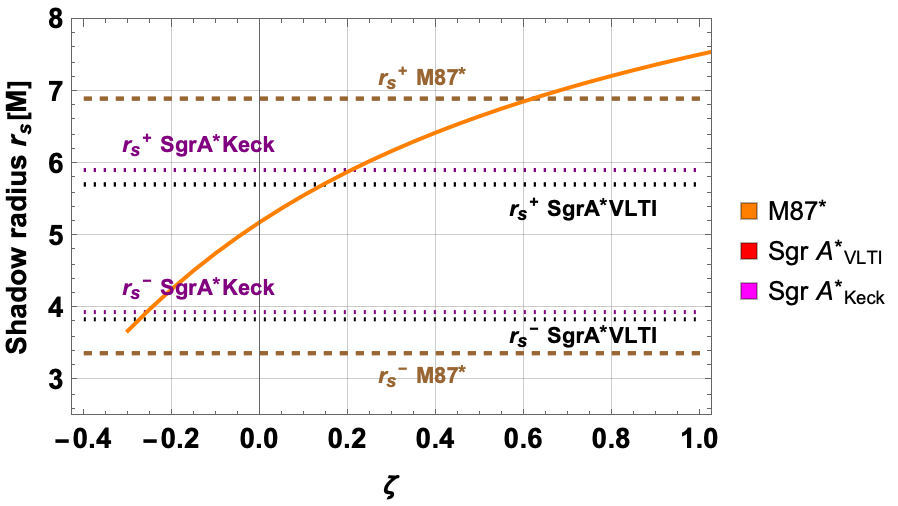}
\end{tabbing}
\caption{\label{rs_bounds} \footnotesize   \it
Shadow radius for Dunkl black hole with M87$^*$ and SgrA$^*$ observed data, {\bf Left:} within the 1-$\sigma$ and {\bf Right:}  the 2-$\sigma$ bounds.}
\end{figure}

\begin{figure}[!ht]
\centering 
			\begin{tabbing}
			\centering
			\hspace{9cm}\=\kill
\includegraphics[width=.5\textwidth]{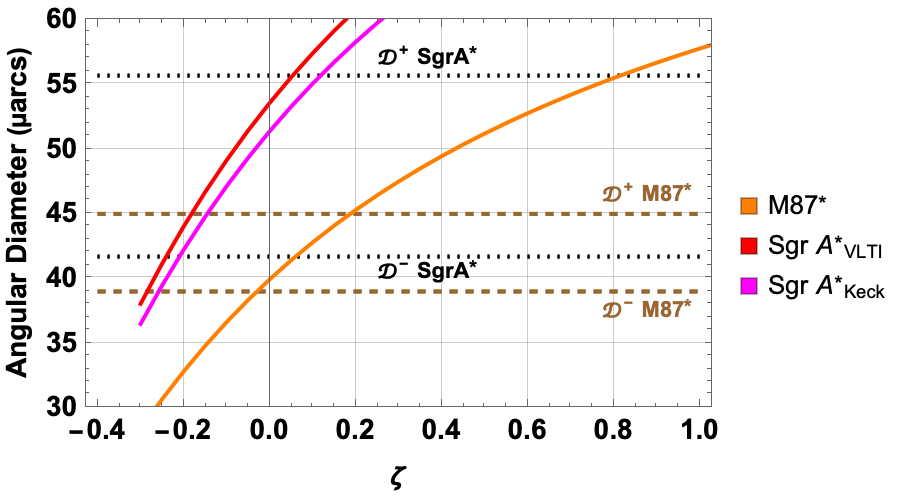}\>
\includegraphics[width=.55\textwidth]{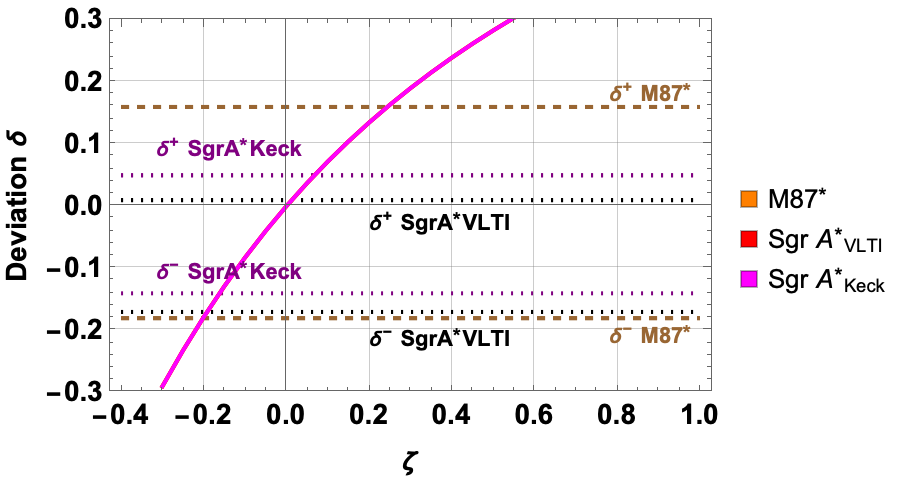}
\end{tabbing}

\caption{\label{angularanddeviation} \footnotesize   \it
Angular diameter $\mathcal{D}$ and fractional deviation $\bm{\delta}$ in terms of the  Dunkl parameter associated with M87$^*$ and SgrA$^*$ black hole observed data.}
\end{figure}

Fig.\ref{rs_bounds}  illustrates  the shadow radius $r_{\text{sh}}/M$ of  the Dunkl black hole  compared with the EHT  shadow size of M87$^*$ within   1-$\sigma$ and 2-$\sigma$ bonds  by varying the Dunkl parameter.  
It is evident that the curves overlap due to the rescaling of mass, with distinctions arising from the specific experimental bounds associated with each black hole. In this sense,  Fig.\ref{rs_bounds}  involves  acceptable   and excluded regions, being compatible and  incompatible with the EHT observations. It gives information on the  ranges of    $\zeta$  in accordance with  EHT collaborations. Within the M87$^*$ data, in the 1-$\sigma$ confidence level, the range  $-0.2043\leq\zeta\leq0.2432$  generates  consistent results.  This range has been enlarged in the  2-$\sigma$ confidence level.   The range  $\zeta> 0.6191 $, however,  provides discordant results with respect to EHT. While for  SgrA$^*$,  we have  found the $-0.194\leq\zeta\leq0.0122$, $-0.2756\leq\zeta\leq0.1444$ within the Keck data  and  $-0.163\leq\zeta \leq 0.0679$,  $-0.2574\leq\zeta \leq 0.270$ for the VLTI measurements, respectively. 
Going further, we   would like to constraint the Dunkl parameter using the angular diameter $\mathcal{D}$  and the fractional deviation ${\bm \delta}$   engineered by EHT collaboration from Keck and VLTI observations as depicted in Fig.\ref{angularanddeviation}. % In order to elaborate a comparative discussion need to provide certain moduli space predictions, it should be interested to mention the estimated  $\sigma$ bonds and deviation observable $\delta$ for two models: 

In addition to the previous requirements  on  the Dunkl parameter $\zeta$  elaborated from the shadow radius $r_s/M$, the constraint derived from the angular diameter $\mathcal{D}$ and the fraction deviation $\bm{\delta}$ illustrated in Fig.\ref{angularanddeviation}  are summarized in Tab.\ref{constraintdunkel}.

\begin{table}[!ht]
\centering
\resizebox{1\textwidth}{!}{%
\begin{tabular}{cc|cccc|}
\cline{3-6}
\multicolumn{2}{c|}{\multirow{2}{*}{}} &
  \multicolumn{4}{c|}{Dunkl parameter $\zeta$} \\ \cline{3-6} 
\multicolumn{2}{c|}{} &
  \multicolumn{1}{c|}{\multirow{2}{*}{$\mathcal{D}$}} &
  \multicolumn{1}{c|}{\multirow{2}{*}{$\bm{\delta}$}} &
  \multicolumn{2}{c|}{$r_s/M$} \\ \cline{5-6} 
 &
   &
  \multicolumn{1}{c|}{} &
  \multicolumn{1}{c|}{} &
  \multicolumn{1}{c|}{1-$\sigma$} &
  2-$\sigma$ \\ \hline
\multicolumn{2}{|c|}{M87$^*$} &
  \multicolumn{1}{c|}{[-0.0316\ ,\ -0.0316]} &
  \multicolumn{1}{c|}{[-0.2044\ ,\ 0,2448]} &
  \multicolumn{1}{c|}{[-0.2043\ ,\ 0.2432]} &
  [---\ ,\ 0.6191] \\ \hline
\multicolumn{1}{|c|}{\multirow{2}{*}{SgrA$^*$}} &
  Keck &
  \multicolumn{1}{c|}{[-0.2418\ ,\ 0,0524]} &
  \multicolumn{1}{c|}{[-0.1924\ ,\ 0.0122]} &
  \multicolumn{1}{c|}{[-0.194\ ,\ 0.0122]} &
  [-0.2756\ ,\ 0.1444] \\ \cline{2-6} 
\multicolumn{1}{|c|}{} &
  VLTI &
  \multicolumn{1}{c|}{[-0.2112\ ,\ 0.1174]} &
  \multicolumn{1}{c|}{[-0.1643\ ,\ 0.0663]} &
  \multicolumn{1}{c|}{[-0.163\ ,\ 0.0679]} &
  [-0.2574\ ,\ 0.270] \\ \hline\hline
\end{tabular}%
}\caption{Estimations of the Dunkel parameter $\zeta$ within the observables $\mathcal{D}$, $\bm{\delta}$ and $r_s/M$ for SgrA$^*$ and M87$^*$ black holes}
\label{constraintdunkel}
\end{table}

\newpage
\section{Conclusion}
In this paper, we have  reconsidered the study of the  Dunkl   black hole by  focusing on the optical aspect. It has shown that such a black hole    has been obtained by combining  the Dunkl operator formalism and the Einstein  gravity equations. The computations  have provided a black hole solution  involving a relevant parameter  refereed to as  $\xi$.  After  a  concise  geometric discussion on such a solution,  we  have investigated  two optical concepts:  the shadow and the deflection angle.   Such notions have been studied by varying the Dunkl parameter. As expected, we have  found  shadow circular geometries since we have considered non-rotating solutions.   The size of the obtained  shadows depends on  the Dunkl parameter.   We have revealed   that the  circular shadow  size  augments   with  the parameter $\xi$.   Considering  small values of the impact parameter,  we have observed  that the deflection angle of light rays  decreases with   such a parameter.    However, the deflection angle of  the light rays   becomes   an   increasing function  for  large values of the impact parameter.

Using   M87$^*$ and SgrA$^*$  bands, we have    provided strong constraints on the Dunkl parameter via  the falsification mechanism by making use of the shadow and the light deflection  angle computations.

This work provides  many open questions.  A natural one  is to add extra parameters enlarging the  Dunkl  black hole moduli space.   We expect deformed  one dimensional curves  with different shapes including  the  cardioid and the elliptic ones.  A  challenging work could be addressed  by introducing Lie algebras via their root systems on which  the Dunkl operators  act in an elegant way.    We leave these  open questions for future investigations.

%\newpage
\bibliographystyle{unsrt}
\bibliography{dunkel_BH_shad.bib}

\end{document}